\documentclass[aps,prx,superscriptaddress,twocolumn,longbibliography]{revtex4-2}
\usepackage{units}
\usepackage{amsmath}
\usepackage{amsthm}
\usepackage{amssymb}
\usepackage{graphicx}
\usepackage{color}
\usepackage{xcolor}
\usepackage{bbold}

\usepackage[colorlinks=true, urlcolor=blue, citecolor=blue,linkcolor=blue,citebordercolor={1 0 0},linkbordercolor={0 0 1}]{hyperref}

\usepackage[linesnumbered, ruled,vlined]{algorithm2e}

\graphicspath{{./Images/},{./ImagesAppendix/}}

\renewcommand{\eqref}[1]{Eq.~(\ref{#1})} 

\theoremstyle{plain}

\theoremstyle{plain}

\ifx\proof\undefined
\newenvironment{proof}[1][\protect\proofname]{\par
	\normalfont\topsep6\p@\@plus6\p@\relax
	\trivlist
	\itemindent\parindent
	\item[\hskip\labelsep\scshape #1]\ignorespaces
}{%
	\endtrivlist\@endpefalse
}
\providecommand{\proofname}{Proof}
\fi
\theoremstyle{plain}

\theoremstyle{remark}

\newcommand{\bra}[1]{\langle #1|}
\newcommand{\ket}[1]{|#1 \rangle}
\makeatother
\newcommand{\braket}[2]{\langle #1 \vert #2 \rangle}
\newcommand{\abs}[1]{\left|#1\right|}
\newcommand{\idg}[1]{{\bfseries #1)}}

\providecommand{\factname}{Fact}
\providecommand{\theoremname}{Theorem}
\providecommand{\claimname}{Claim}
\providecommand{\lemmaname}{Lemma}
\providecommand{\definitionname}{Definition}

\definecolor{KB}{rgb}{0.4,0.3,0.9}

\definecolor{THc}{rgb}{0.9,0.3,0.2}

\newcommand{\revA}[1]{{#1}}
\newcommand{\revB}[1]{{#1}}
\newcommand{\revC}[1]{{#1}}
\theoremstyle{definition}

\newcommand{\subfigimg}[3][,]{%
	\setbox1=\hbox{\includegraphics[#1]{#3}}
	\leavevmode\rlap{\usebox1}
	\rlap{\hspace*{2pt}\raisebox{\dimexpr\ht1-0.5\baselineskip}{{\bfseries \large\textsf{#2}}}}
	\phantom{\usebox1}
}

\begin{document}
\title{Capacity and quantum geometry of parametrized quantum circuits}
\author{Tobias Haug}
\email{thaug@ic.ac.uk}
\affiliation{QOLS, Blackett Laboratory, Imperial College London SW7 2AZ, UK}
\author{Kishor Bharti}
\affiliation{Centre for Quantum Technologies, National University of Singapore 117543, Singapore}
\author{M. S. Kim}
\affiliation{QOLS, Blackett Laboratory, Imperial College London SW7 2AZ, UK}
\begin{abstract}
To harness the potential of noisy intermediate-scale quantum devices, it is paramount to find the best type of circuits to run hybrid quantum-classical algorithms. Key candidates are parametrized quantum circuits that can be effectively implemented on current devices. 
Here, we evaluate the capacity and trainability of these circuits using the geometric structure of the parameter space via the effective quantum dimension, which reveals the expressive power of circuits in general as well as of particular initialization strategies.
We assess the expressive power of various popular circuit types and find striking differences depending on the type of entangling gates used. Particular circuits are characterized by scaling laws in their expressiveness. 
We identify a transition in the quantum geometry of the parameter space, which leads to a decay of the quantum natural gradient for deep circuits. For shallow circuits, the quantum natural gradient can be orders of magnitude larger in value compared to the regular gradient; however, both of them can suffer from vanishing gradients.
By tuning a fixed set of circuit parameters to randomized ones, we find a region where the circuit is expressive, but does not suffer from barren plateaus, hinting at a good way to initialize circuits. We show an algorithm that prunes redundant parameters of a circuit without affecting its effective dimension.
Our results enhance the understanding of parametrized quantum circuits and can be immediately applied to improve variational quantum algorithms.
\end{abstract}
\maketitle

\section{Introduction}

Quantum computers promise to tackle challenging problems for classical computers such as drug design, combinatorial optimisation and simulation of many-body physics. 
While fully-fledged large-scale quantum computers with error correction are not expected to be available for many years, noisy intermediate-scale quantum (NISQ) devices have been investigated as a way to approach computationally hard problems with quantum processors available now and in the near future~\cite{preskill2018quantum,bharti2021noisy}. 
Variational quantum algorithms (VQA)~\cite{peruzzo2014variational,mcclean2016theory,cerezo2020variational,cao2019quantum} have been a major hope in achieving a quantum speedup with NISQ devices. 
The core idea is to update a parametrized quantum circuit (PQC) in a hybrid quantum-classical fashion. Measurements performed on the PQC are fed into a classical computer to propose a new set of variational parameters.
A key challenge has been the occurrence of barren plateaus, i.e. the gradients used for optimisation vanish exponentially with increasing number of qubits~\cite{mcclean2018barren}, as well as for various types of cost functions~\cite{cerezo2021cost}, entanglement~\cite{marrero2020entanglement} and noise~\cite{wang2020noise}. Further, the classical optimization part of variational algorithms was shown to be NP-hard~\cite{bittel2021training}. 
Quantum algorithms that avoid the feed-back loop to circumvent the barren plateau problems have been proposed~\cite{huang2019near,bharti2020quantum,bharti2020quantum2,bharti2020iterative,haug2020generalized,lau2021quantum,lim2021fast}.
Besides this approach, initialization strategies~\cite{volkoff2020large,grant2019initialization,grimsley2019adaptive} and layer-wise learning~\cite{skolik2020layerwise} for VQA could help to solve the aforementioned problems. However, tools to evaluate the power of these strategies are lacking.
Hardware efficient ans\"atze have been proposed to tailor a PQC to the restrictions of the hardware~\cite{kandala2017hardware}. A widely used choice is quantum circuits arranged in layers of single-qubit rotations followed by two-qubit entangling gates. 
However, a key question is the space of possible states this ansatz type can express~\cite{nakaji2020expressibility,sim2019expressibility,rasmussen2020reducing,funcke2020dimensional}.

Here, we introduce the effective quantum dimension $G_\text{C}$ and parameter dimension $D_\text{C}$ as a quantitative measure of the capacity of a PQC. 
Parameter dimension $D_\text{C}$ measures the total number of independent parameters a quantum state defined by the PQC can express. In contrast, the effective quantum dimension $G_\text{C}$~\cite{mackay1992bayesian,maddox2020rethinking} is a local measure to quantify the space of states that can be accessed by locally perturbing the parameters of the PQC. Both measures can be derived from the quantum geometric structure of the PQC via the quantum Fisher information metric (QFI) $\mathcal{F}$~\cite{yamamoto2019natural,stokes2020quantum}. 
From the QFI, one can obtain the quantum natural gradient (QNG) for a more efficient optimisation via gradients~\cite{yamamoto2019natural,stokes2020quantum,wierichs2020avoiding}.
These methods allow us to evaluate the expressive power, trainability and number of redundant parameters of different PQCs, and find better initialization strategies.  

\begin{figure*}[htbp]
	\centering
	\includegraphics[width=0.8\textwidth]{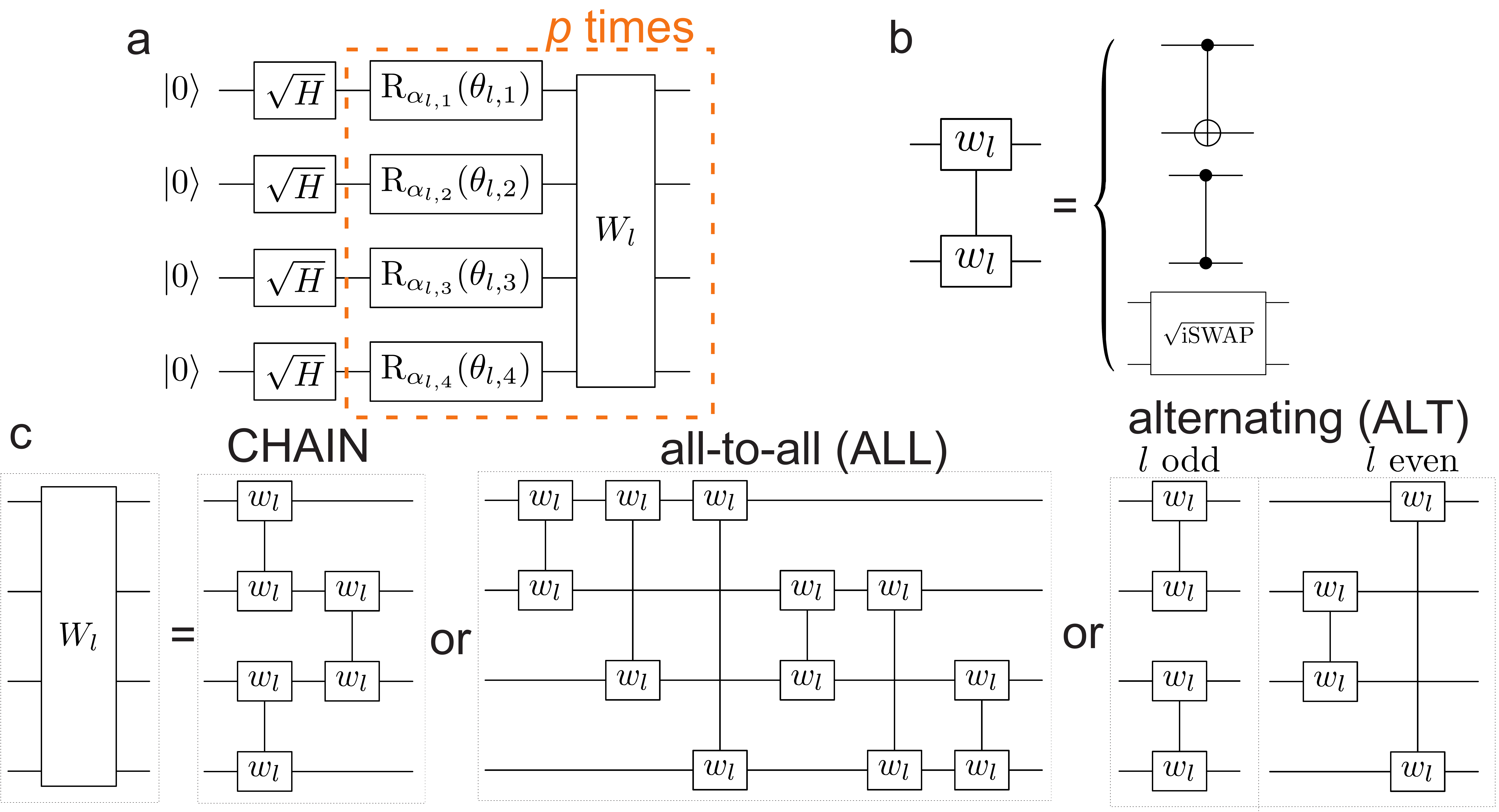}
	\caption{\idg{a} Sketch of hardware-efficient parametrized quantum circuit (PQC) $U(\theta)\ket{0}^{\otimes N}=\prod_{l=p}^1 \left[  W_l V_l(\theta_l)\right]\sqrt{H_\text{d}}^{\otimes N}\ket{0}^{\otimes N}$ with parameters $\theta$ and initial state $\ket{0}$ of all $N$ qubits being in state zero. The PQC consists of an initial layer of $\sqrt{H_\text{d}}$ gates applied to each qubit, where $H_\text{d}$ is the Hadamard gate, followed by $p$ repeated layers of parametrized single qubit rotations $V_l(\theta_l)$ and entangling gates $W_l$. $V_l(\theta_l)$ consists of single-qubit rotations $R_\alpha(\theta_{l,n})=\exp(-i\sigma^\alpha_n\theta_{l,n}/2)$ at layer $l$ and qubit $n$ around axis $\alpha\in\{x,y,z\}$. 
	\idg{b} Two-qubit entangling gates $w_l$ considered are CNOT gates (control-$\sigma_x$), CPHASE  gates (control-$\sigma_z$, $\text{diag}(1,1,1,-1)$) or $\sqrt{\text{iSWAP}}$ gates.
	\idg{c} Entangling layer $W_l$ is composed of the two-qubit entangling gates $w_l$, which are arranged in either a nearest-neighbor one-dimensional chain topology (denoted as CHAIN), all-to-all connection (ALL) or in a alternating fashion (ALT) for even and odd layers $l$.}
	\label{SketchCircuit}
\end{figure*}

As demonstration of our tools, we provide an in-depth investigation of popular hardware-efficient circuits, composed of layered single-qubit rotations and two-qubit entangling gates in various arrangements. We find striking differences depending on the choice of circuit structure that affect both the expressive power of the PQC in general as well as the quality of specific initialization strategies.
We calculate the number of redundant parameters of various PQC types, as well as how fast they converge towards random quantum states as a function of the number of layers. 
The choice of entangling gate has a pronounced effect on the expressive power of particular initialization strategies.

We reveal a transition in the spectrum of the QFI in deep circuits, which leads to a decay of the QNG.
For shallow circuits, the QNG can be orders of magnitude larger than the regular gradient. However, both suffer from the barren plateau problem.
By tuning the PQCs parameters from zero to a random set of parameters, we find a region where both large gradients and large effective quantum dimension $G_\text{C}$ coexist, which could serve as a good set of initial parameters for the training of variational algorithms. 
\revA{Finally, as an application of our method we propose and apply an algorithm that prunes redundant parameters from PQCs, while keeping the parameter dimension constant. This algorithm helps us to find expressive PQCs with a reduced number of parameters to simplify training for variational algorithms as well as a reduced circuit depth to ease the impact of noise.}

\revB{The paper is organized as follows. First, we define PQCs in Sec.\ref{sec:paramcircuit} and the parameter dimension $D_\text{C}$ in Sec.\ref{sec:paramdimension}. Then, we introduce the effective quantum dimension $G_\text{C}$ in Sec.\ref{sec:effdimension}.
Our results and the algorithm are presented in Sec.\ref{sec:results}, which are discussed in Sec.\ref{sec:discussion}.
We give an overview of the definitions of symbols in Tab.\ref{tab:definitions}.}

\begin{table*}[htbp]\centering
\begin{tabular}{ |c|c|c| } 
\hline
Name&Symbol & Definition  \\
\hline
Parameterized quantum circuit (PQC) & $\ket{\psi(\theta)}$&$U(\theta)\ket{0}^{\otimes N}$\\
Quantum Fisher information metric (QFI) &$\mathcal{F}_{ij}(\theta)$&$\text{Re}(\braket{\partial_i \psi}{\partial_j \psi}-\braket{\partial_i \psi}{\psi}\braket{\psi}{\partial_j \psi})$\\
Expectation value of Hamiltonian $H$  &$E$&$E=\bra{0}U^\dagger(\theta)HU(\theta)\ket{0}$\\
Quantum natural gradient &QNG&$\mathcal{F}^{-1}(\theta)\partial_k E$\\
Variance & $\text{var}(E)$&$\langle (E)^2\rangle -\langle E \rangle^2$\\
Effective dimension&$G_\text{C}(\theta)$ & $\text{rank}(\mathcal{F}(\boldsymbol{\theta}))$ \\ 
Parameter dimension&$D_\text{C}$ & $G_\text{C}(\theta_\text{random})$ \\ 
Number of parameters of a PQC &$M$&\\
Redundancy&$R$&$(M-D_\text{C})/M$\\
\hline
\end{tabular}
\caption{Definitions of symbols.}
\label{tab:definitions}
\end{table*}

\section{Parametrized quantum circuits}\label{sec:paramcircuit}
A PQC generates a quantum state of $N$ qubits  
\begin{equation}
\ket{\psi(\theta)}=U(\theta)\ket{0}^{\otimes N}\,,
\end{equation}
with the unitary $U(\theta)$, the $M$-dimensional parameter vector $\theta$ and product state $\ket{0}^{\otimes N}$ as shown in Fig.\ref{SketchCircuit}.
The structure of the PQC influences its power to express quantum states~\cite{nakaji2020expressibility,sim2019expressibility,du2020expressive}. 
One way to measure expressiveness is by determining the distance between the distribution of states generated by the circuit and the Haar random distribution of states~\cite{nakaji2020expressibility,sim2019expressibility}. This tells us how well the PQC can express arbitrary states across the Hilbert space. 
The appearance of barren plateaus or vanishing gradients is connected to the aforementioned measure~\cite{mcclean2018barren,holmes2021connecting}. 
The variance of the gradient $\text{var}(\partial_i E)=\langle (\partial_i E)^2\rangle -\langle \partial_i E \rangle^2$ ($\langle .\rangle$ denoting statistical average over many random instances) in respect to the expectation value of a Hamiltonian $H$ ($E=\bra{0}U^\dagger(\theta)HU(\theta)\ket{0}$) can vanish exponentially with the number of qubits for PQCs with a random choice of parameters. 
The variance decreases also with number of layers $p$ of the PQC until a specific $p_\text{r}$, where it remains constant upon further increase of $p>p_\text{r}$. For local cost functions, it has been shown that in most cases low variance of the gradient of such PQCs correlates with high expressibility~\cite{holmes2021connecting}.

\section{Parameter dimension}\label{sec:paramdimension}
We now introduce the parameter dimension $D_{\text{C}}$ of a PQC as another measure of capacity. As example, we take a PQC \revB{that can represent arbitrary $N$ qubit quantum states} which is parametrized by in total $M=2^{N+1}$ parameters $\boldsymbol{a}$, $\boldsymbol{b}$
\begin{equation}
\ket{\psi(\boldsymbol{a},\boldsymbol{b})}=\sum_{j=1}^{2^N}(a_j+ib_j)\ket{j}\, ,\label{eq:state}
\end{equation}
where $\ket{j}$ is the $j$-th computational basis state and $a_j, b_j\in \mathcal{R}$. 
One can map the above state to ${D_\text{C}=2^{N+1}-2}$ independent parameters, that lie on the surface of ${2^{N+1}-1}$ dimensional sphere. \revB{Of the in total $M$ parameters, the final $2$ parameters are dependent and do not change the quantum state, as they correspond to the norm and global phase of the quantum state.}
Conversely, for a generic real-valued quantum state with $b_j=0$, we find ${D_{\text{C}}=2^{N}-1}$ independent parameters, \revB{with $1$ dependent parameter due to the norm of the real-valued quantum state}.
Analogous to the generic quantum state, we now define the parameter dimension $D_\text{C}$ for a PQC $C$ as the number of independent parameters that the PQC can express in the space of quantum states.
In general,  $D_\text{C}$ for $N$ qubits is upper bounded by the generic state~\eqref{eq:state} \revB{with ${D_\text{C}\le 2^{N+1}-2}$.}
We define the redundancy
\begin{equation}
R=\frac{M-D_\text{C}}{M}\,,\label{eq:redundancy}
\end{equation}
which is the fraction of dependent parameters of the PQC that do not contribute to changing the quantum state.
\revB{In the next section, we show how $D_\text{C}$ can be determined for hardware efficient PQCs.}

\begin{figure}[htbp]
	\centering
	\includegraphics[width=0.27\textwidth]{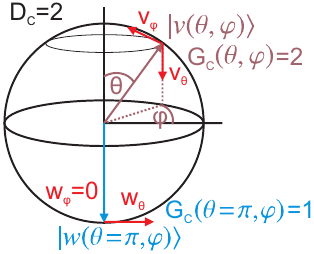}
	\caption{Example to demonstrate the effective quantum dimension $G_\text{C}$ and parameter dimension $D_\text{C}$ for a single qubit parametrized as $\ket{\psi(\theta,\varphi)}=\cos(\theta/2)\ket{0}+\exp(i\varphi)\sin(\theta/2)\ket{1}$. \revB{The quantum state is described by $M=2$ parameters $\theta$ and $\varphi$.} 
	$D_\text{C}=2$ is the number of \emph{independent} parameters of the quantum state. $G_\text{C}(\theta, \varphi)$ denotes the number of independent directions the quantum state can change to by locally perturbing its parameters $\theta$, $\varphi$. For a random state $\ket{v}$ ($\theta\notin\{0,\pi\}$) two possible directions exist, along $v_\theta$ and $v_\varphi$. The particular state $\ket{w(\theta=\pi,\varphi)}$ can only be perturbed in direction $w_\theta$ as adjusting $\varphi$ does not change the state (e.g. $\ket{w(\pi,\varphi+\epsilon)}=\ket{w(\pi,\varphi)})$, thus $G_\text{C}(\theta=\pi,\varphi)=1$.  }
	\label{SketchBloch}
\end{figure}

\section{Effective quantum dimension}\label{sec:effdimension}

Now, we explain how the QFI $\mathcal{F}(\theta)$ quantifies the expressive power of a PQC (see Appendix~\ref{sec:Fisherinfo} for an introduction to the QFI and QNG, and Appendix~\ref{sec:meas_qfi} on how to calculate it).
One can relate $\mathcal{F}(\theta)$ to the distance in the space of pure quantum states, which is given by the Fubini-Study distance
\begin{equation}\label{eq:fdist}
    {\rm Dist}_Q\Big(\ket{\psi(\theta)}, ~ \ket{\psi(\theta+\text{d}\theta)} \Big)^2
      = \sum_{i,j}\mathcal{F}_{ij}(\theta)\text{d}\theta_i \text{d}\theta_j\,, 
\end{equation}
where ${\rm Dist}_Q(x,y)=\vert\braket{x}{y}\vert^2$ and the QFI~\cite{stokes2020quantum,yamamoto2019natural}
\begin{equation}\label{eq:QFI}
\mathcal{F}_{ij}(\theta)=\text{Re}(\braket{\partial_i \psi}{\partial_j \psi}-\braket{\partial_i \psi}{\psi}\braket{\psi}{\partial_j \psi})\,,
\end{equation}
which corresponds to the real part of the quantum geometric tensor.
$\mathcal{F}(\theta)$ quantifies the change of the quantum state when adjusting its parameter $\theta$ infinitesimally to $\theta+\text{d}\theta$. 
The eigenvalue decomposition
\begin{equation}\label{eq:SVD}
\mathcal{F}(\theta)=V S V^T\,,
\end{equation}
gives us $V$, which is a real-valued unitary with the $i$-th eigenvector $\alpha^{(i)}$ placed at the $i$-th column of $V$, and $S$, which is a diagonal matrix with the $M$ non-negative eigenvalues $\lambda^{(i)}$ of $\mathcal{F}(\theta)$ along the diagonal. The eigenvalues and eigenvectors obey the equation $\mathcal{F}(\theta)\alpha^{(i)}=\lambda^{(i)}\alpha^{(i)}$. 
Inserting \eqref{eq:SVD} into \eqref{eq:fdist} gives us
\begin{equation}\label{eq:fdistEig}
    {\rm Dist}_Q\Big(\ket{\psi(\theta)}, ~ \ket{\psi(\theta+d\theta)} \Big)
      = \text{d}\theta^T \mathcal{F} d\theta=d\theta^T V S V^T \text{d}\theta \,. 
\end{equation}
Now, we assume that the small variations in $\theta$ are in the direction of the $i$-th eigenvector of $\mathcal{F}(\theta)$ with $\text{d}\theta=\text{d}\mu \alpha^{(i)}$, where $\text{d}\mu$ is an infinitesimal scalar. We find
\begin{align*}
    {\rm Dist}_Q&\Big(\ket{\psi(\theta)}, ~ \ket{\psi(\theta+d\mu \alpha^{(i)})} \Big)^2
      = \\
      &d\mu {\alpha^{(i)}}^T V S V^T \alpha^{(i)}\text{d}\mu=\lambda^{(i)}\text{d}\mu \text{d}\mu \,,
\end{align*}
where we have used $V^T\alpha^{(i)}=e^{(i)}$, where $e^{(i)}$ is the $i$-th basis vector.
When updating $\theta'=\theta+\text{d}\mu\alpha^{(i)}$, the quantum state changes at a rate that is proportional to $\lambda^{(i)}$.
Eigenvalues $\lambda^{(i)}=0$ are called singularities as there is no change in the quantum state at all, i.e. $\vert\braket{\psi(\theta)}{\psi(\theta+\text{d}\mu \alpha^{(i)})}\vert=1$. 
The case $\lambda^{(i)}$ being very small, i.e. $1\gg\lambda^{(i)}>0$, is called near singularity and is associated with plateaus in classical machine learning where training slows down~\cite{amari2016information}.

We now define the effective quantum dimension $G_\text{C}(\theta)$ for a PQC $C$ as the rank of the QFI $\mathcal{F}(\theta)$. It is given as the total number of non-zero eigenvalues $\lambda^{(i)}(\theta)$ of  $\mathcal{F}(\theta)$ initialized with parameters $\theta$~\cite{mackay1992bayesian,maddox2020rethinking}
\begin{equation}
    G_\text{C}(\theta)=\sum_{i=1}^M \mathcal{I}(\lambda^{(i)}(\theta))\, ,
\end{equation}
where $\mathcal{I}(x)=0$ for $x=0$ and $\mathcal{I}(x)=1$ for $x\neq0$. 
$G_\text{C}(\theta)$ is a local measure of expressiveness that counts the number of independent directions in the state space that can be accessed by an infinitesimal update of $\theta$.

A straightforward example is a generic single qubit quantum state shown in Fig.\ref{SketchBloch}
\begin{gather}
\ket{\psi(\theta,\varphi)}=\cos\left(\frac{\theta}{2}\right)\ket{0}+\exp(i\varphi)\sin\left(\frac{\theta}{2}\right)\ket{1}\\
\mathcal{F}(\theta)= \left[\begin{array}{cc}
        1 & 0 \\
        0 & \sin^2(\theta) \\
    \end{array}\right]. 
\end{gather}
The eigenvalues and eigenvectors of the QFI $\mathcal{F}$ are straightforward to calculate
with $\lambda_1=1$, $\alpha_1=\{1,0\}$ and $\lambda_2=\sin^2(\theta)$, $\alpha_2=\{0,1\}$.
The effective quantum dimension is $G_\text{C}(\theta,\varphi)=D_\text{C}=2$, except for the special case $\theta=n\pi$, $n$ integer, wjere the eigenvalue is $\lambda_2=0$ and thus $G_\text{C}(n\pi,\varphi)=1$. Here, any change in the direction of eigenvector $\alpha_2$ (corresponding to changing $\varphi$) will not yield any change in the underlying quantum state. 
However note that except for these singular parameters we find $G_\text{C}=2$, which is equivalent to the maximal number of independent parameters $D_C$ of the system.

As further example we consider the single qubit circuit with Pauli $z$ matrix $\sigma_z$ and Hadamard gate $H_\text{d}$
\begin{equation}
U(\theta)\ket{0}=\prod_{i=1}^M\left[\exp(-i\frac{\theta_i}{2}\sigma_z)\right]H_\text{d}\ket{0}\,.
\end{equation}
Here, we find $\mathcal{F}=\frac{1}{4}J_{M,M}$, where $J_{M,M}$ is a $M\times M$ matrix filled with ones. Diagonalizing $\mathcal{F}$ gives us $M-1$ eigenvalues with $\lambda=0$, and one eigenvalue $\lambda_1=\frac{M}{4}$ with eigenvector $\alpha_1=\frac{1}{\sqrt{M}}J_{M,1}$. This circuit has a low parameter dimension $D_\text{C}=1$ and a large redundancy of $R=(M-1)/M$, i.e. there are $M-1$ parameter directions which do not yield any change of the quantum state.

For the type of PQC as shown in Fig.\ref{SketchCircuit}, which are arranged in a layer-wise structure with the parametrized gates being Pauli operators,  the effective quantum dimension $G_\text{C}$ is equal or less than the parameter dimension $D_\text{C}$, which in turn is equal or less than the number of parameters $M$
\begin{equation}
    G_\text{C}(\theta)\le D_\text{C}\le M\,.
\end{equation}
Given the aforementioned PQC types with a random set of parameters $\theta_\text{random}\in\text{random}(0,2\pi)$, we find numeric evidence that $G_\text{C}(\theta_\text{random})$ is approximately equivalent to $D_\text{C}$
\begin{equation}
    G_\text{C}(\theta_\text{random})\simeq D_\text{C}\,.\label{eq:Grandom}
\end{equation}
Thus, we can calculate $D_\text{C}$ by determining $G_\text{C}(\theta_\text{random})$ for random sets of PQC parameters.
The core intuition is that starting from a sufficiently random initial parameter set, a change of the PQC parameters in the right direction is able to bring one closer to any quantum state that can be expressed by the PQC.
For specific choices of parameters such as $\theta=0$ we find $G_\text{C}< D_\text{C}$. Moving sufficiently away from these special points, we recover that $G_\text{C}\simeq D_\text{C}$.

We stress that \eqref{eq:Grandom} is not valid for arbitrary quantum circuits, e.g. circuits where the parameters do not enjoy a $2\pi$ periodicity. As simple example take the evolution of a single qubit with a single parameter $t$ $U(t)\ket{0}=\exp(-i\sqrt{2}\sigma_z t)\exp(-i \sigma_x t)\ket{0}$. The evolution over all possible $t$ (note the absence of $2\pi$ periodicity) will cover all possible quantum states and thus $D_\text{C}=2$, whereas the effective quantum dimension (with only a single parameter $t$) is $G_\text{C}=1< D_\text{C}$.

\begin{figure}[htbp]
	\centering
	\subfigimg[width=0.24\textwidth]{a}{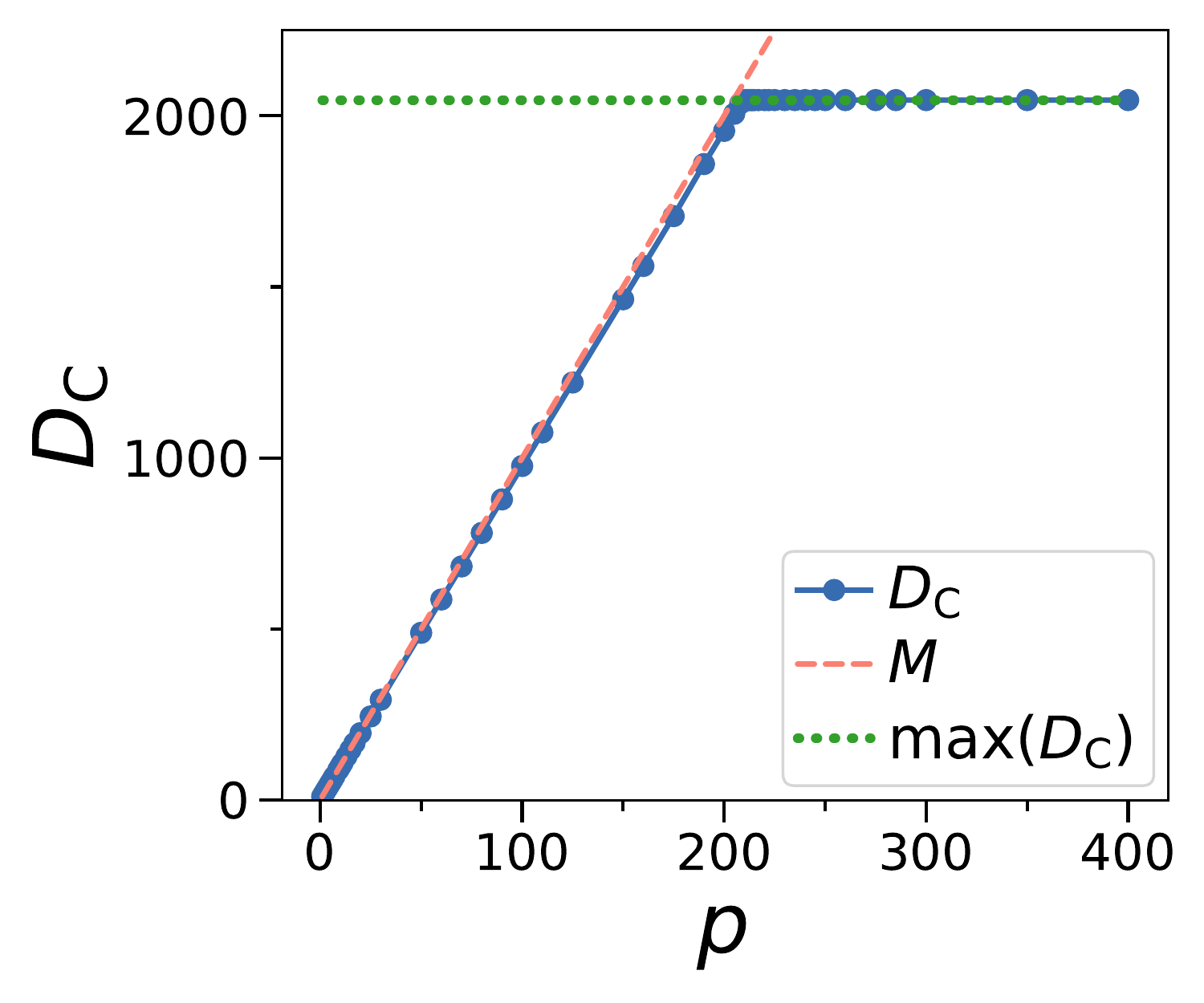}\hfill
	\subfigimg[width=0.24\textwidth]{b}{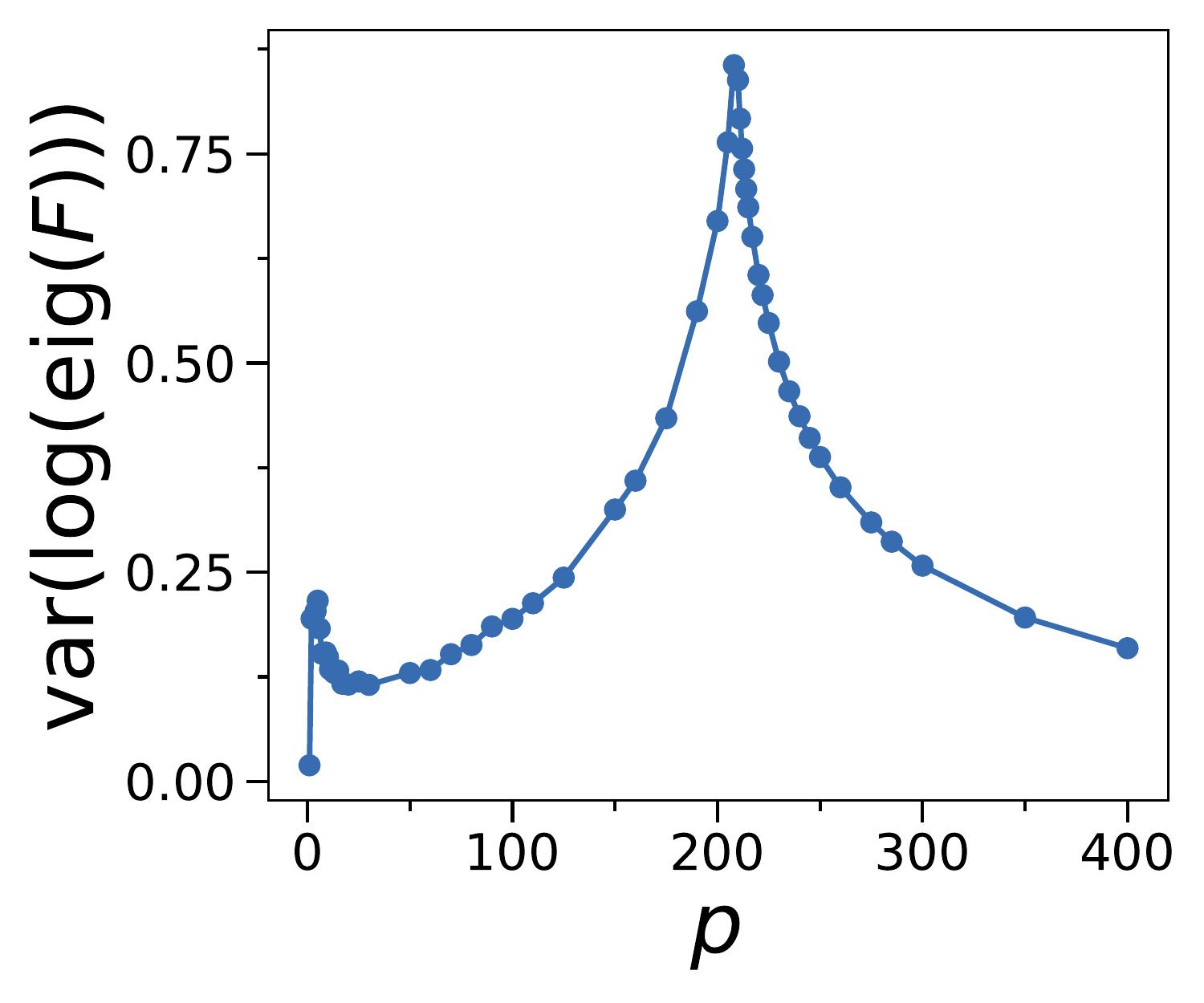}\\
	\subfigimg[width=0.24\textwidth]{c}{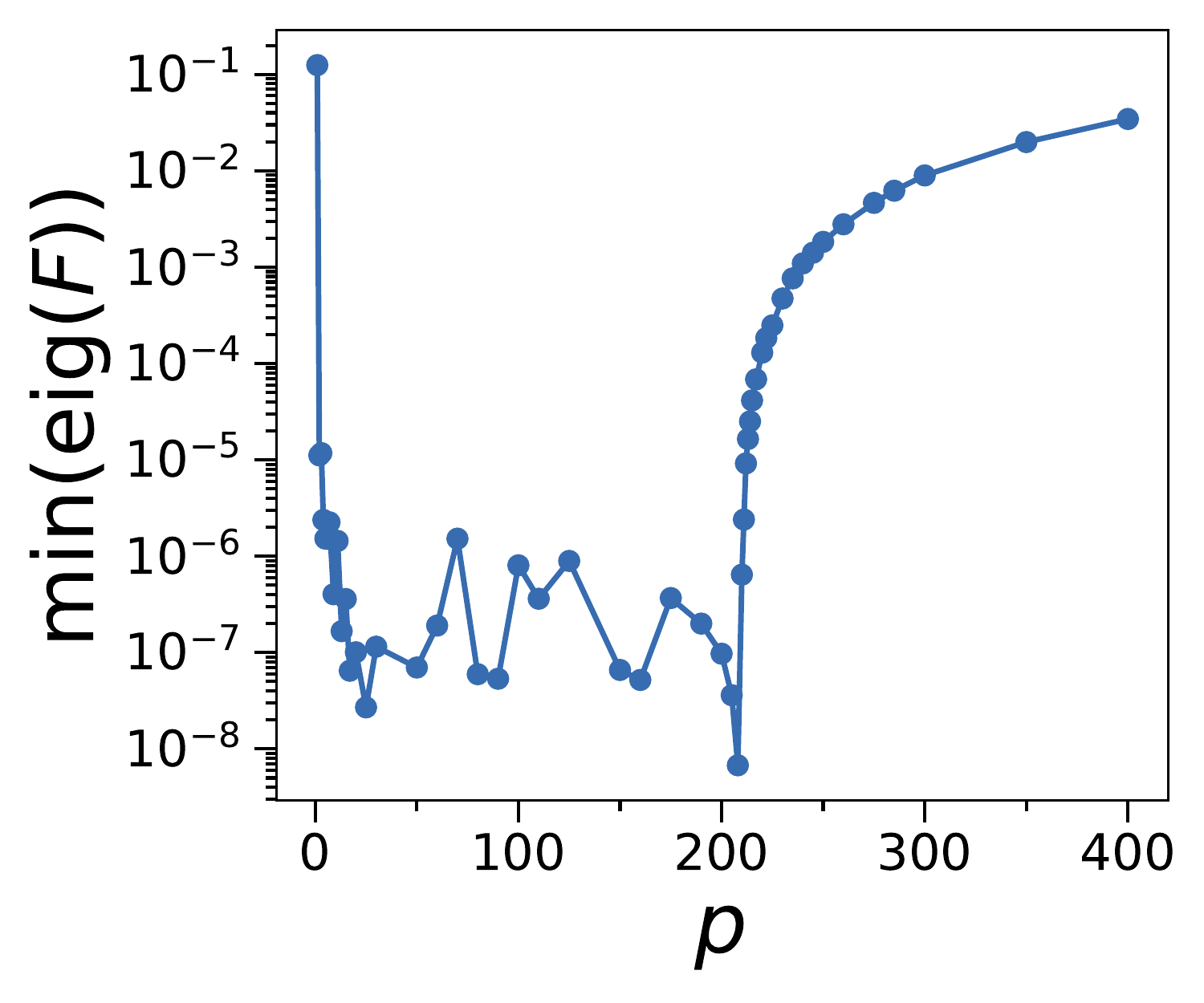}\hfill
	\subfigimg[width=0.24\textwidth]{d}{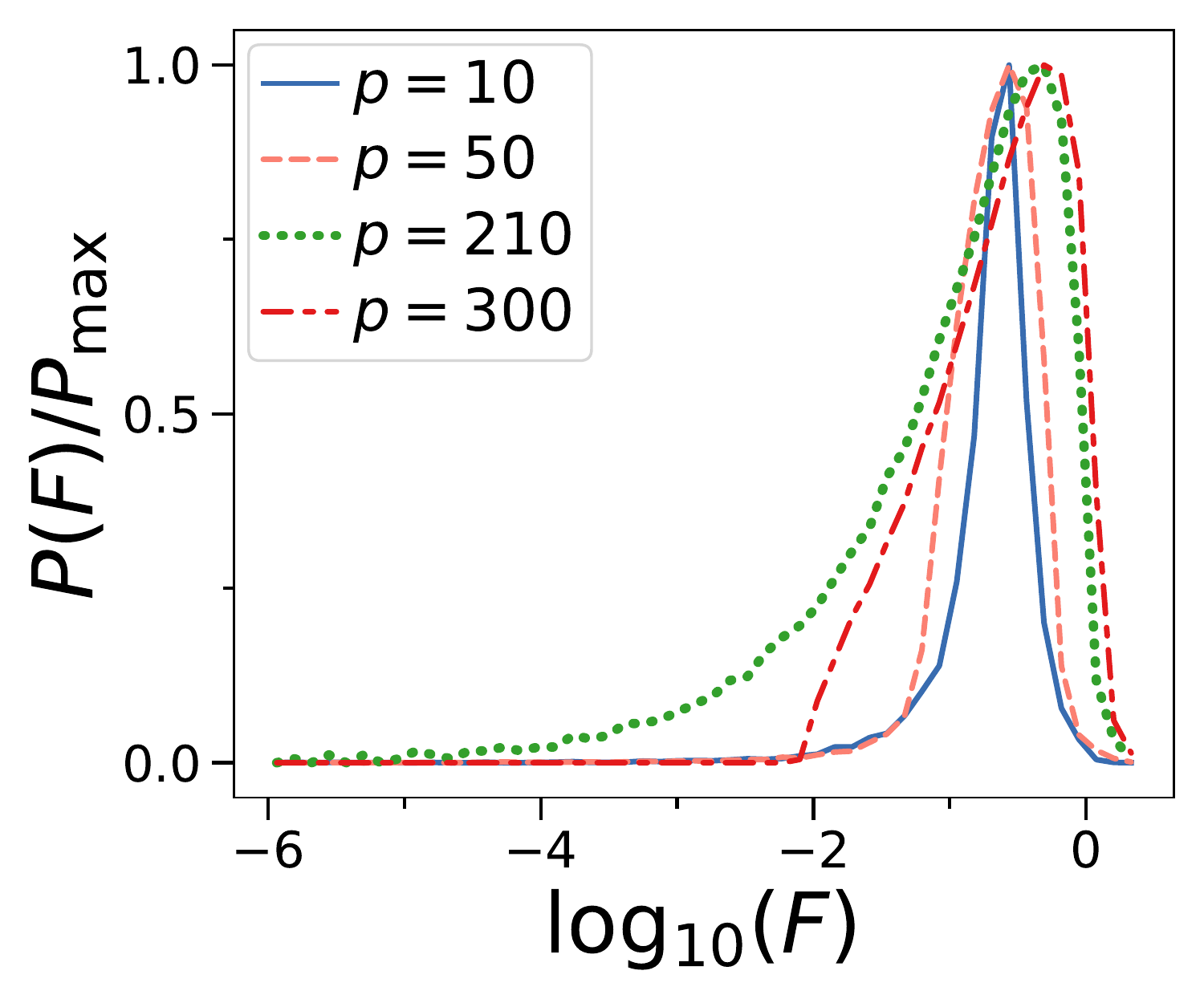}\\
	\subfigimg[width=0.24\textwidth]{e}{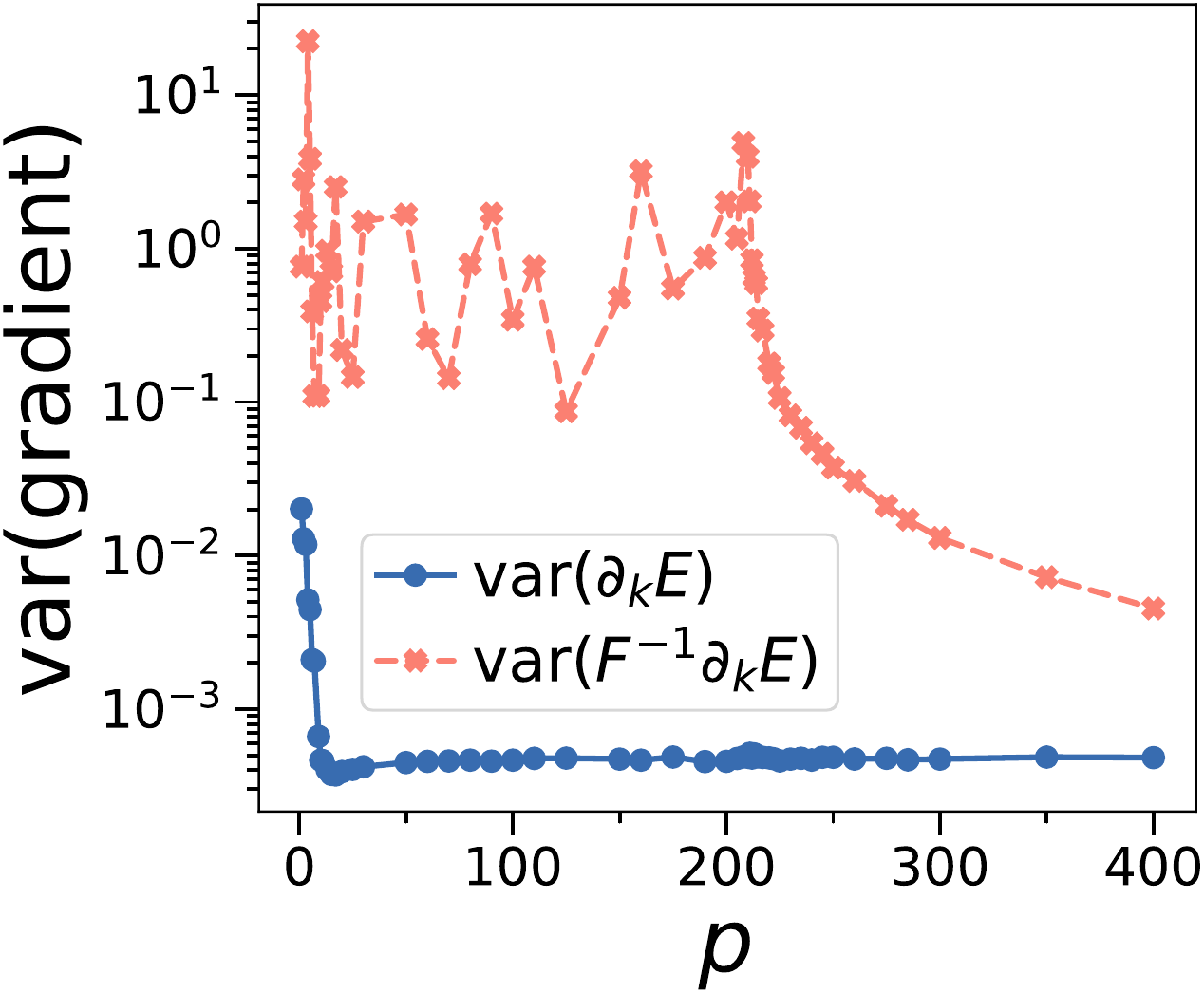}\hfill
	\subfigimg[width=0.24\textwidth]{f}{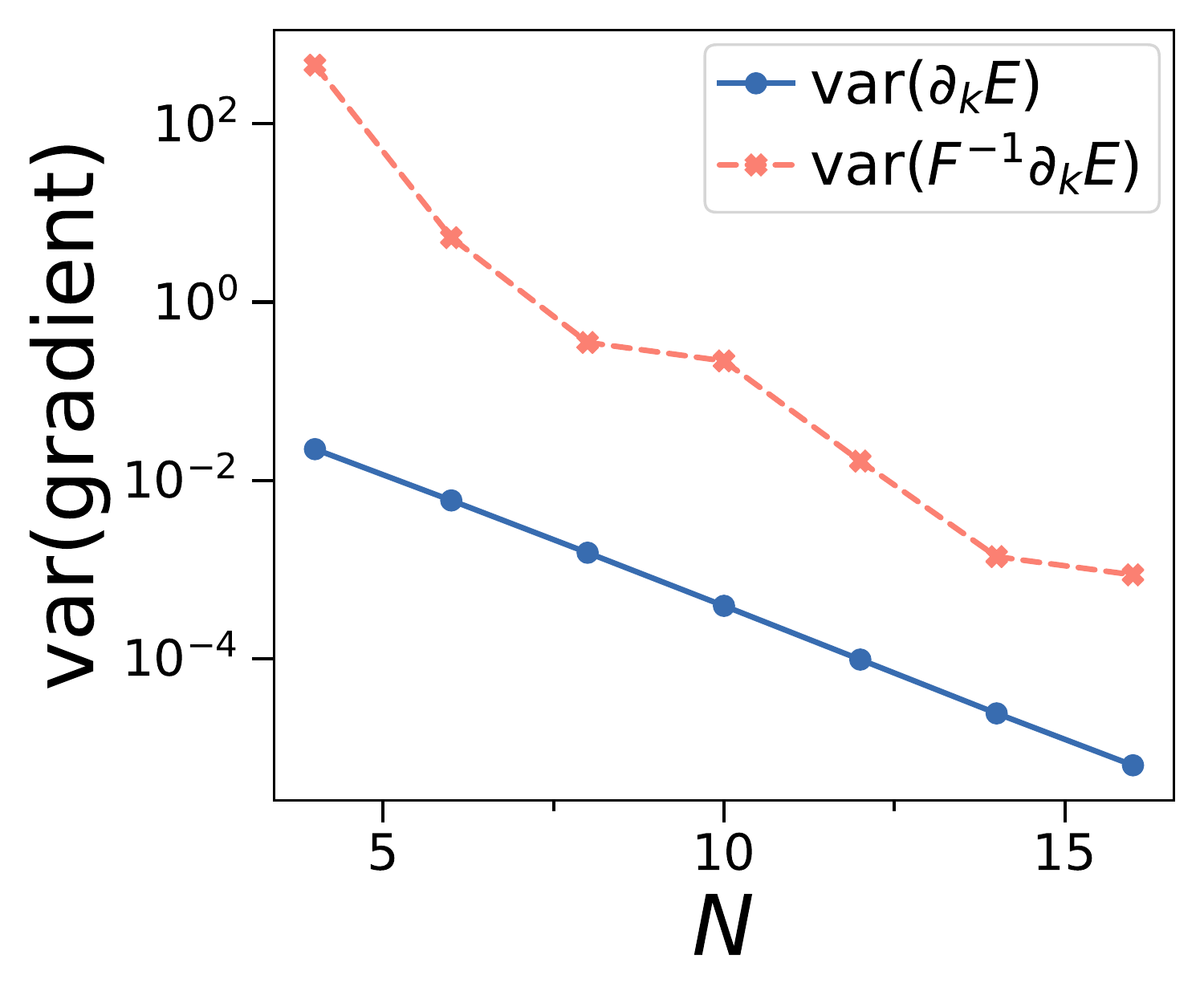}
	\caption{Properties of PQC consisting of $p$ layers of randomly chosen $x$, $y$, $z$ rotations, followed by CNOT gates in a chain topology (see Fig.\ref{SketchCircuit}) for $N=10$ qubits. 
	\idg{a} The parameter dimension $D_\text{C}$ of the PQC (determined via Eq.\ref{eq:Grandom}) scales linearly with $p$, until it levels at a characteristic value $p_\text{c}\approx210$.
	\idg{b} Variance of the logarithm of the non-zero eigenvalues of $\mathcal{F}$. The variance peaks around $p\approx p_\text{c}$. 
	\idg{c} Minimal non-zero eigenvalue of $\mathcal{F}$ against $p$. It increases for $p>p_\text{c}$. 
	\idg{d} Histogram of logarithm of eigenvalues of Fisher information matrix $\mathcal{F}$. The width of the distribution increases with $p$, with a pronounced tail at small $\mathcal{F}$ developing around $p\approx p_\text{c}$, which disappears for $p>p_\text{c}$.
	\idg{e} Variance of the gradient $\text{var}(\partial_k E)$ and QNG $\text{var}(\mathcal{F}^{-1}\partial_k E)$ in respect to the Hamiltonian $H=\sigma_1^z\sigma_2^z$. The gradient decays until $p\approx 20$, after which it remains constant. The QNG remains larger than the regular gradient, but decreases for $p>p_\text{c}$. 
	\idg{f} Variance of gradients and QNG for varying qubit number $N$ for depth $p=2N$, showing approximate exponential decrease with $N$.
	}
	\label{CNOT}
\end{figure}

We now consider different types of hardware efficient PQC $\ket{\psi(\theta)}=U(\theta)\ket{0}^{\otimes N}$, which are circuits that can be efficiently run on NISQ quantum processors. We choose an initial state $\ket{0}^{\otimes N}$, followed by a single layer of the square root of the Hadamard gate ($\sqrt{H_\text{d}}$) on every qubit. Then, we repeat $p$ layers composed of parametrized single qubit rotations and a set of two-qubit entangling gates (see Fig.\ref{SketchCircuit}a). The single qubit rotations are either chosen randomly to be around the $\{x,y,z\}$ axis, or fixed to a specific axis. 
The two-qubit entangling gates are either CNOT, CPHASE or $\sqrt{\text{iSWAP}}$ gates (see Fig.\ref{SketchCircuit}b), that are common native gates in current quantum processors~\cite{krantz2019quantum}.
The entangling gates in each layer are arranged in either a nearest-neighbor chain topology (CHAIN), all-to-all connections (ALL) or in an alternating nearest-neighbor fashion (ALT) (see Fig.\ref{SketchCircuit}c). 
The numerical calculations are performed using Yao~\cite{yao}.

\section{Results}\label{sec:results}
As a demonstration of our methods, we provide an in-depth characterization of a PQC consisting of randomly chosen $x$, $y$, $z$ rotations and CNOT gates in a chain topology as function of number of layers $p$ in Fig.\ref{CNOT}. 
The parameter dimension $D_\text{C}$ (i.e. number of independent parameters of the quantum state that can be expressed by the PQC) increases linearly with $p$ in Fig.\ref{CNOT}a, until it reaches the maximal possible value for $D_\text{C}=2^{N+1}-2$ at a characteristic number of layers $p_\text{c}$. 
This point is reflected in the spectrum of the QFI $\mathcal{F}$, averaged over random instances of the PQC (see Fig.\ref{CNOT}b-d). Most notably, the variance of the logarithm of the non-zero eigenvalues  reaches a maximum for $p_\text{c}$ (Fig.\ref{CNOT}b). Further, the minimum taken over all eigenvalues becomes minimal (Fig.\ref{CNOT}c). We can see this more clearly in the distribution of eigenvalues (Fig.\ref{CNOT}d). With increasing $p$, the distribution becomes broader, with a pronounced tail of small eigenvalues of $\mathcal{F}$ appearing close to the transition at $p_\text{c}$. Above the transition $p>p_\text{c}$, the small eigenvalues suddenly disappear from the distribution. 
We investigate the variance of the gradient and QNG in Fig.\ref{CNOT}e \revB{for the two-qubit Hamiltonian $H=\sigma_1^z\sigma_2^z$}.  The variance of the regular gradient decays with $p$, reaching a minimum around $p\approx20$~\cite{mcclean2018barren}, upon which it remains constant. The variance of the QNG remains larger than the regular gradient, however the QNG decays for $p>p_\text{c}$. 
In Fig.\ref{CNOT}f, we numerically find that variance of both regular gradient and QNG vanish exponentially with increasing number of qubits $N$, demonstrating the barren plateau problem. \revB{In the Appendix~\ref{sec:ising_decay}, we show that the same result is found also for more complicated Hamiltonians such as the transverse Ising model.}

\begin{figure}[htbp]
	\centering
	\subfigimg[width=0.24\textwidth]{a}{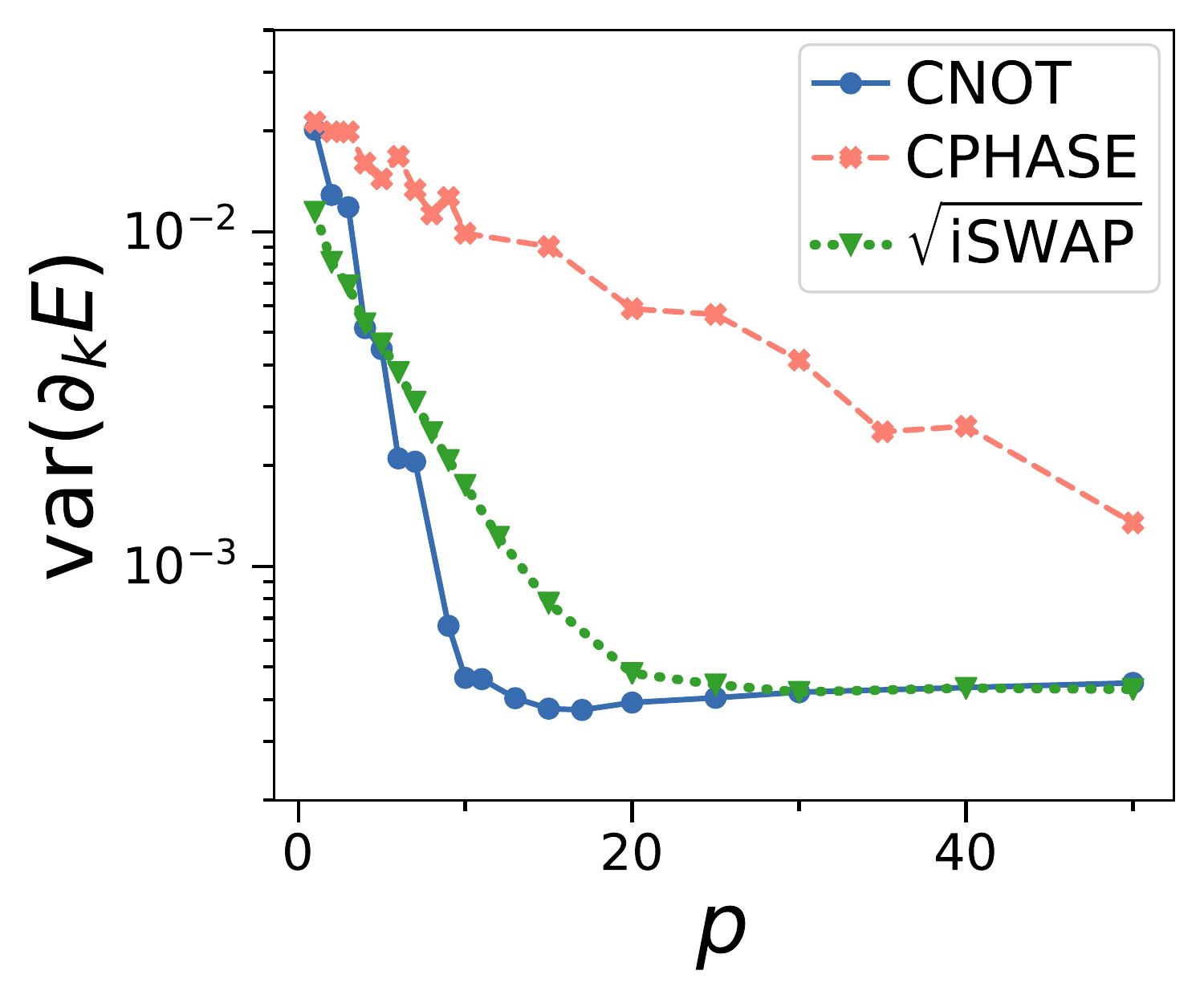}\hfill
	\subfigimg[width=0.24\textwidth]{b}{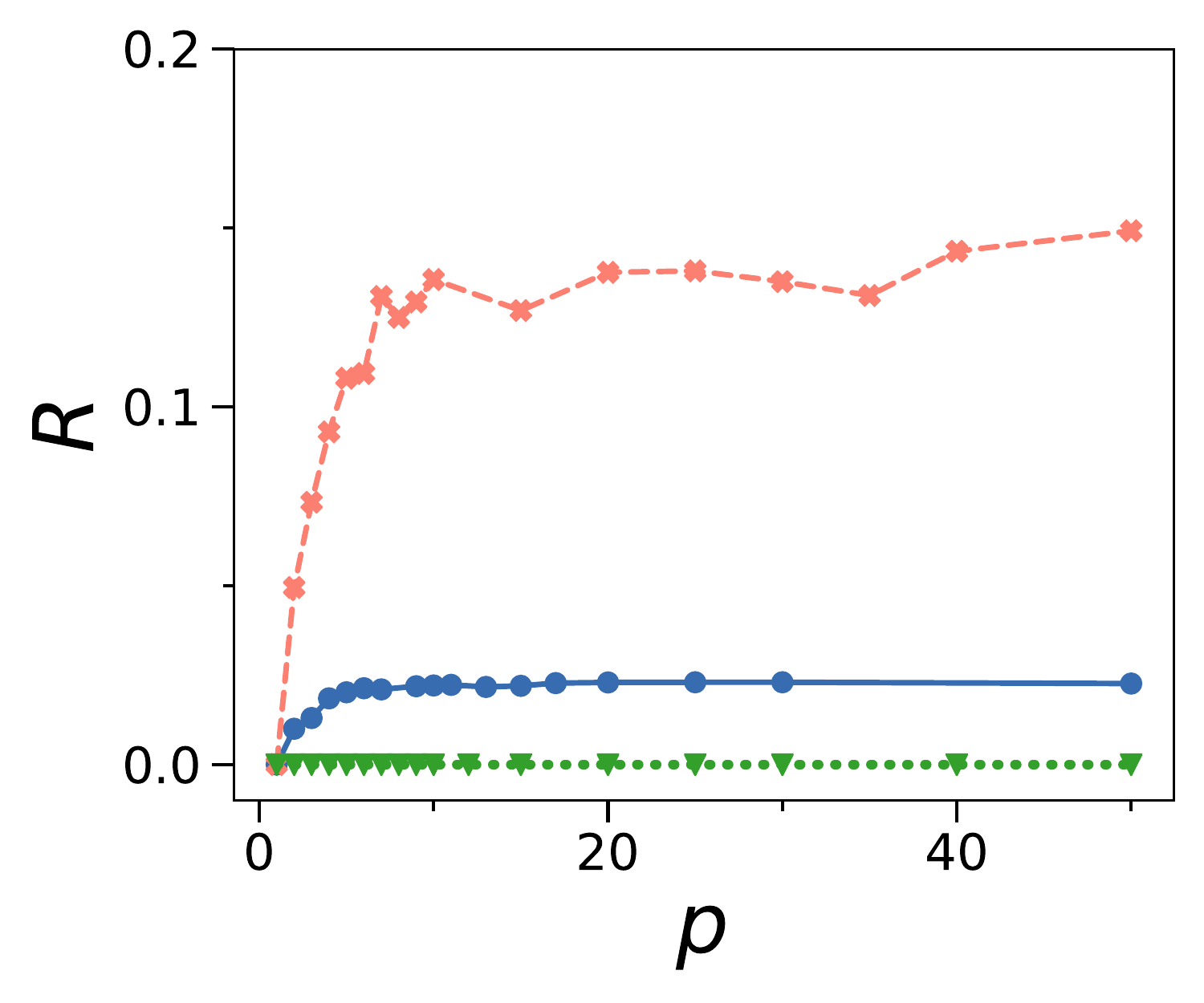}\\
	\subfigimg[width=0.24\textwidth]{c}{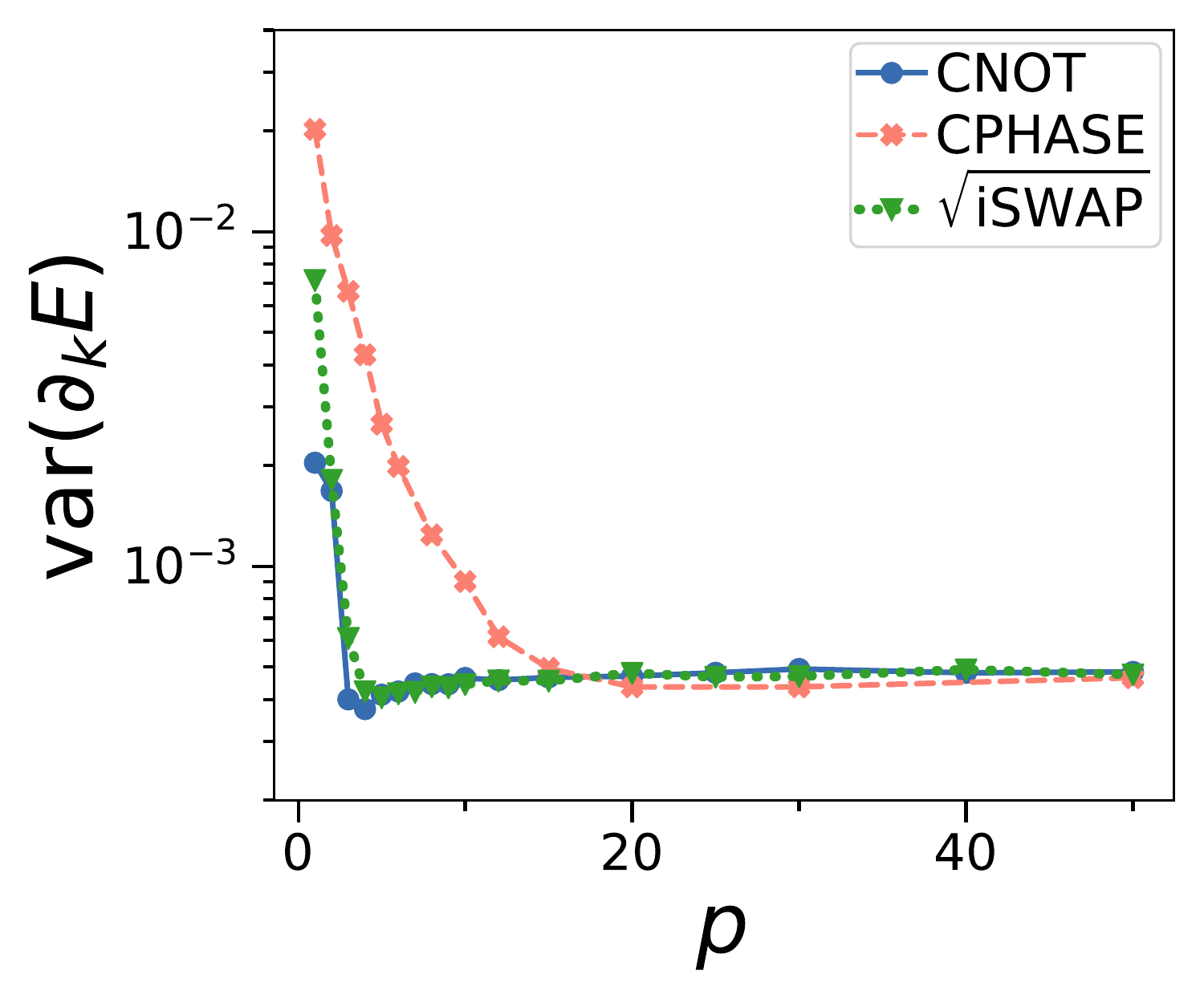}\hfill
	\subfigimg[width=0.24\textwidth]{d}{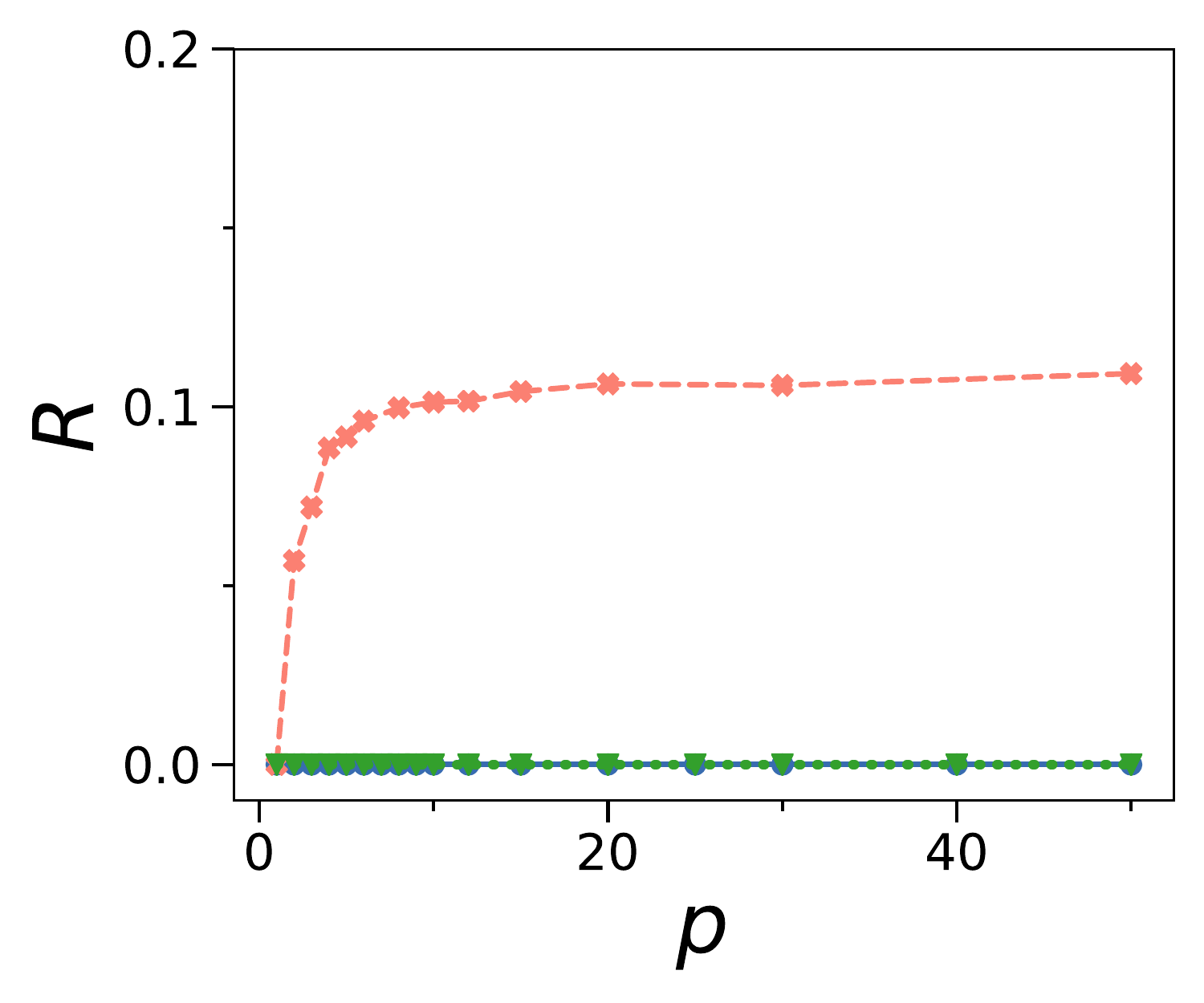}\\
	\subfigimg[width=0.24\textwidth]{e}{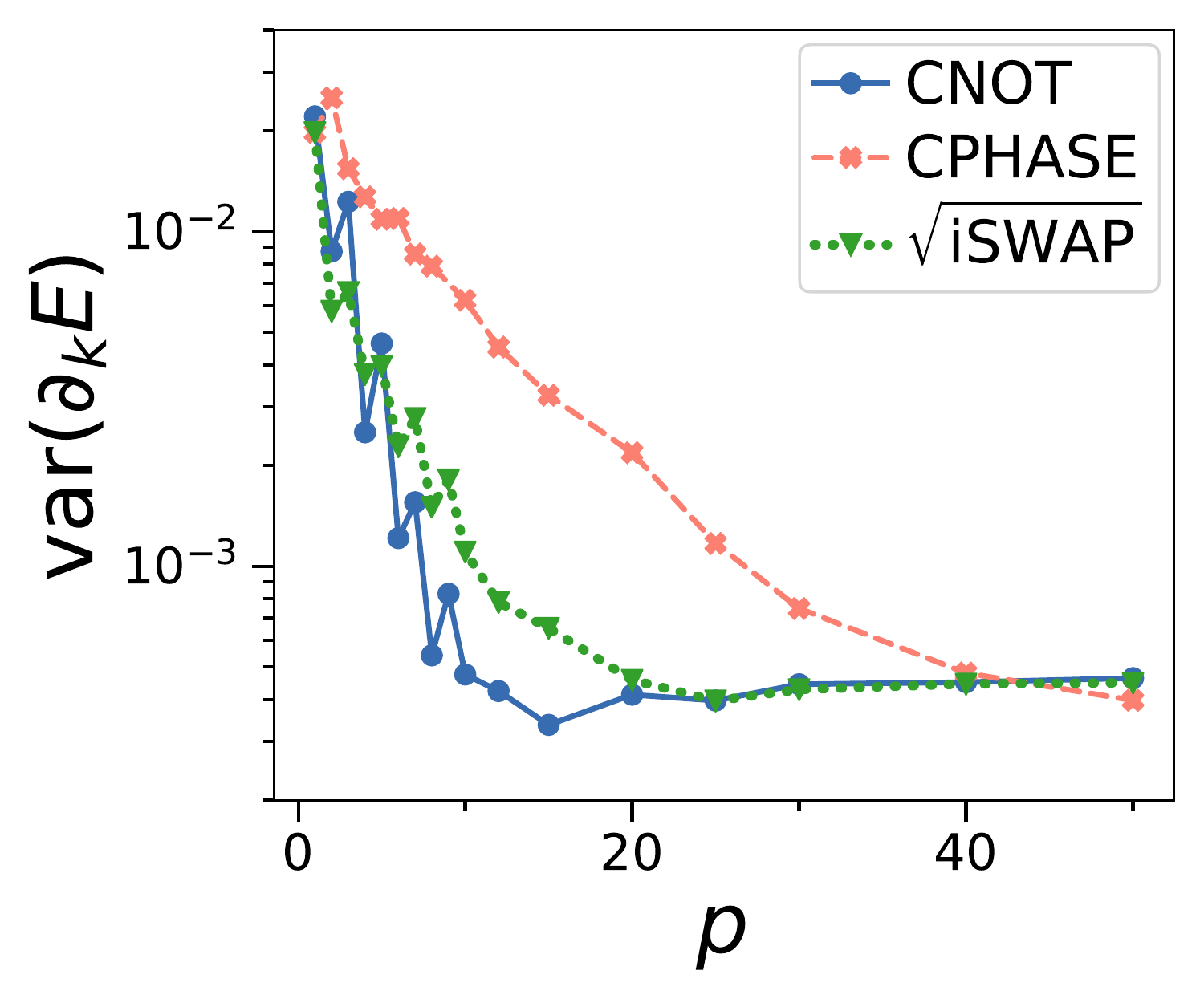}\hfill
	\subfigimg[width=0.24\textwidth]{f}{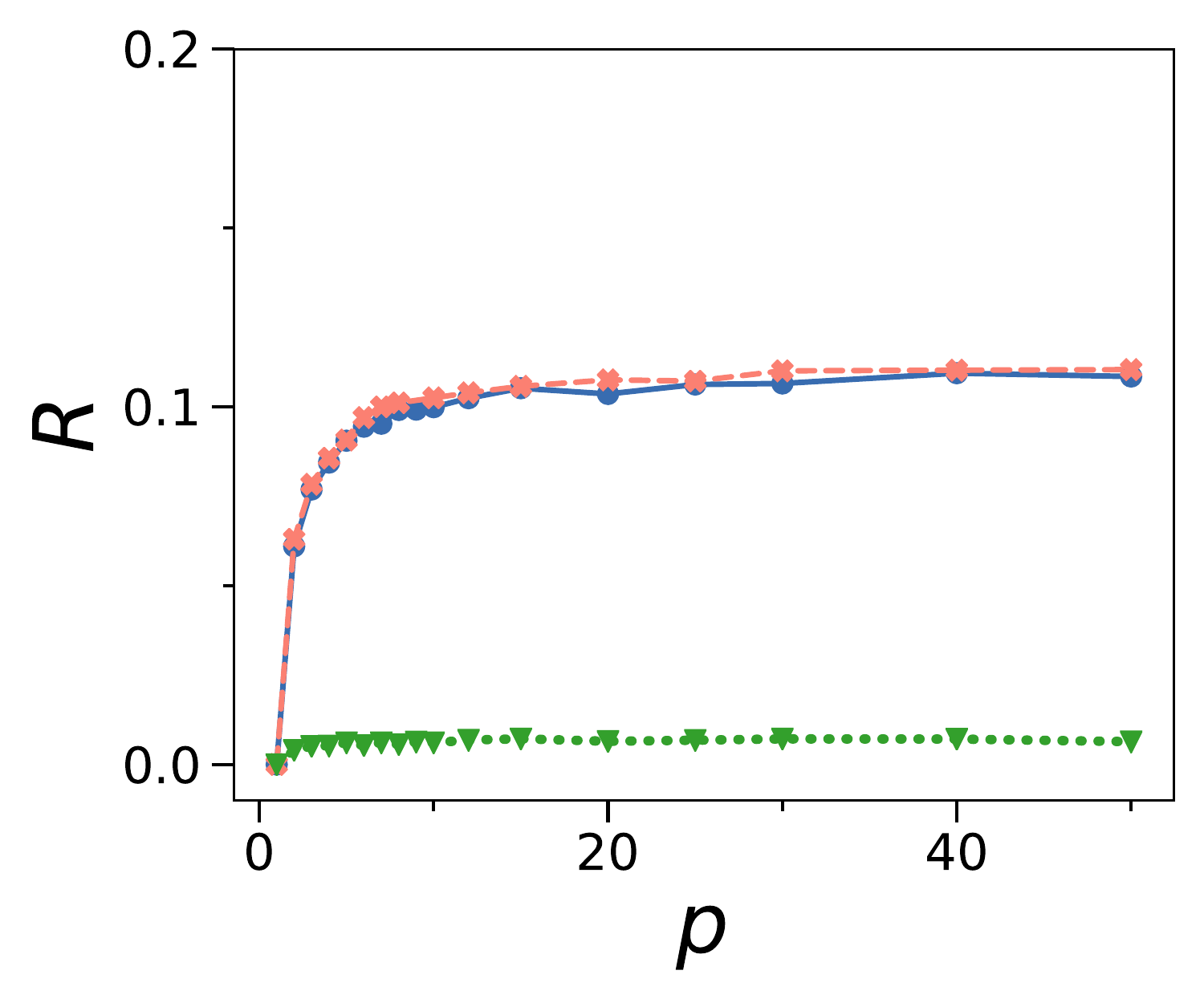}
	\caption{Variance of gradient and redundancy of different hardware efficient PQCs plotted against number of layers $p$ for $N=10$ qubits. Each layer consists of single qubit-rotations as randomly chosen rotations around $\{x,y,z\}$ axis. We plot different arrangements of entangling gates (as shown in Fig.\ref{SketchCircuit}c) with \idg{a,b} nearest-neighbor one-dimensional chain, \idg{c,d} all-to-all, \idg{e,f} alternating nearest-neighbor connections.
	\idg{a,c,e} Variance of the gradient $\text{var}(\partial_k E)$ in respect to the Hamiltonian $H=\sigma_1^z\sigma_2^z$.
	\idg{b,d,f} Redundancy $R$ (\eqref{eq:redundancy}), which is the fraction of redundant parameters of the PQC.}
	\label{CompareCircuits}
\end{figure}

In Fig.\ref{CompareCircuits}, we compare different types of PQCs with different entangling gates and arrangements. 
We note that all circuits show the same qualitative behavior regarding the transition in the QFI (see Fig.\ref{CNOT} and Appendix~\ref{sec:further_data}) as well as suffer from exponential decrease of the variance of the gradient with increasing number of qubits. 
However, key differences in the different PQCs appear.
We show the variance of the gradient \revB{for the Hamiltonian $H=\sigma_1^z\sigma_2^z$} in Fig.\ref{CompareCircuits}a,c,e for different arrangements of the entangling gates (CHAIN, ALL, ALT) as well as different types of entangling gates (CNOT, CPHASE, $\sqrt{\text{iSWAP}}$). 
The variance decays with increasing $p$, until it reaches a constant level, the value of which is the same for all gates and arrangements. However, CPHASE requires the most layers $p$ to converge, followed by $\sqrt{\text{iSWAP}}$ and CNOT. 
Fig.\ref{CompareCircuits}b,d,f shows the redundancy $R$, which is the fraction of redundant parameters of the PQC. It quickly reaches a constant level with increasing $p$. $\sqrt{\text{iSWAP}}$ has consistently low $R$, while for CNOT it varies depending on the arrangement of entangling gates. For CPHASE, we have consistently larger $R$. This can be easily understood when considering that $z$ rotations commute with the entangling CPHASE layer. When two $z$ rotations appear consecutively on the same qubit, they yield a redundant parameter. $R$ for CNOT depends highly on the entangling gates arrangement.

We note that for these PQCs the number of layers $p_\text{c}$ at which the transition of the QFI occurs can be estimated from the value of redundancy $R$. We find $p_\text{c}\approx(1-R_\text{C})(2^{N+1}-2)/b$, where $R_\text{C}$ is the value of $R$ for sufficiently large $p$ and $b$ is the number of parameterized rotations per layer. 
The eigenvalue spectrum of these PQCs and further types of PQCs are discussed in Appendix~\ref{sec:further_data}.

In Fig.\ref{yRotation}, we fix the single-qubit rotations around the $y$-axis and investigate different entangling gates arranged in a nearest-neighbor one-dimensional chain. Depending on the choice of entangling gates, we find that the variance of the gradient \revB{for $H=\sigma_1^z\sigma_2^z$} decays to a different constant level with increasing $p$ (see Fig.\ref{yRotation}a). $y$ $\sqrt{\text{iSWAP}}$ matches the variance found in Fig.\ref{CNOT}e, whereas $y$ CNOT and $y$ CPHASE have higher variance.
In Fig.\ref{yRotation}b we show the maximal $D_\text{C}$ for many layers $p$. 
$D_\text{C}$ scales exponentially for $y$ CNOT ($D_\text{C}\propto 2^{N}$) and $y$ $\sqrt{\text{iSWAP}}$ ($D_\text{C}\propto 2^{N+1}$), whereas for $y$ CPHASE we find numerically an approximate quadratic scaling $D_\text{C}\propto N^2$.

\begin{figure}[htbp]
	\centering
	\subfigimg[width=0.24\textwidth]{a}{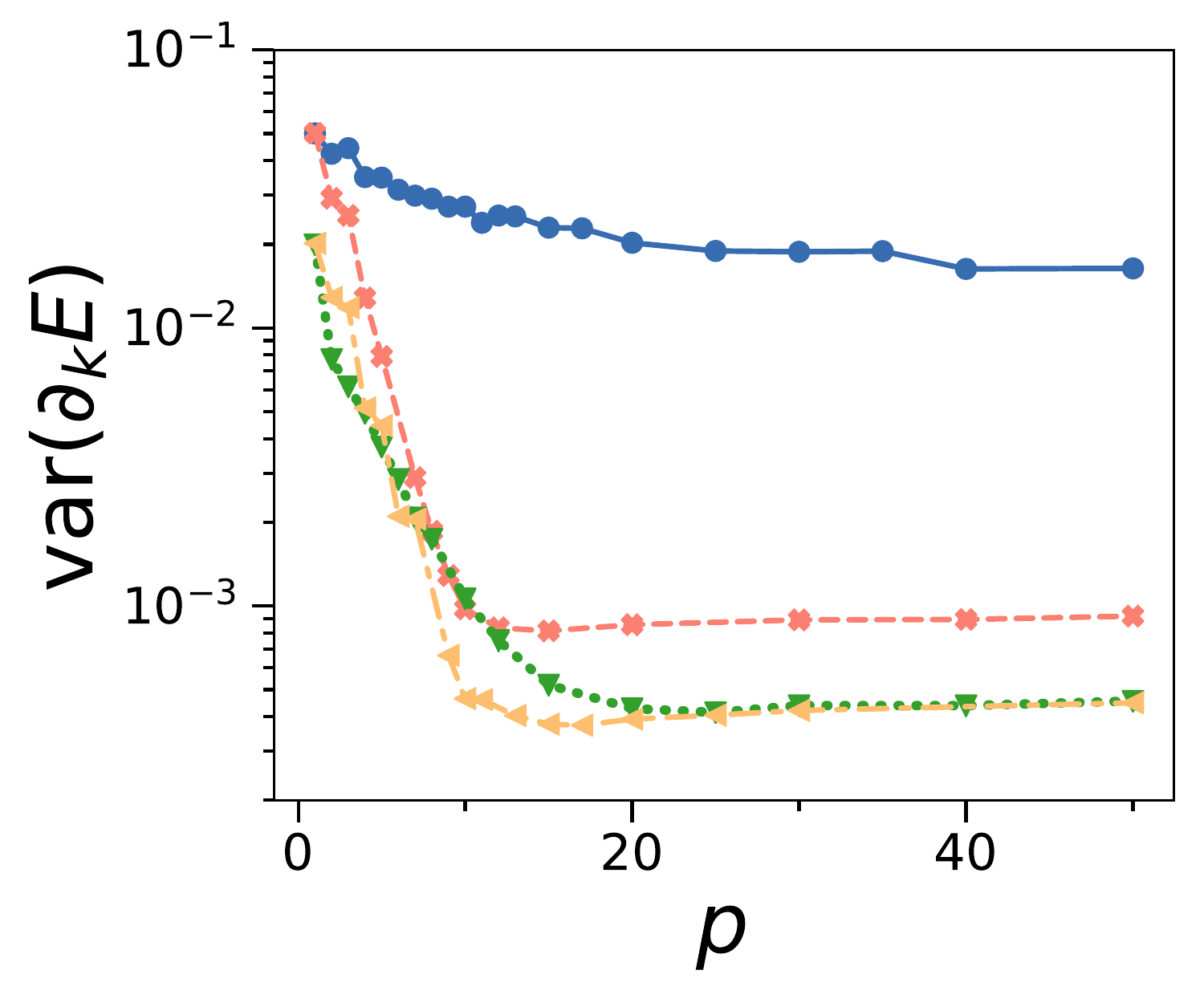}\hfill
	\subfigimg[width=0.24\textwidth]{b}{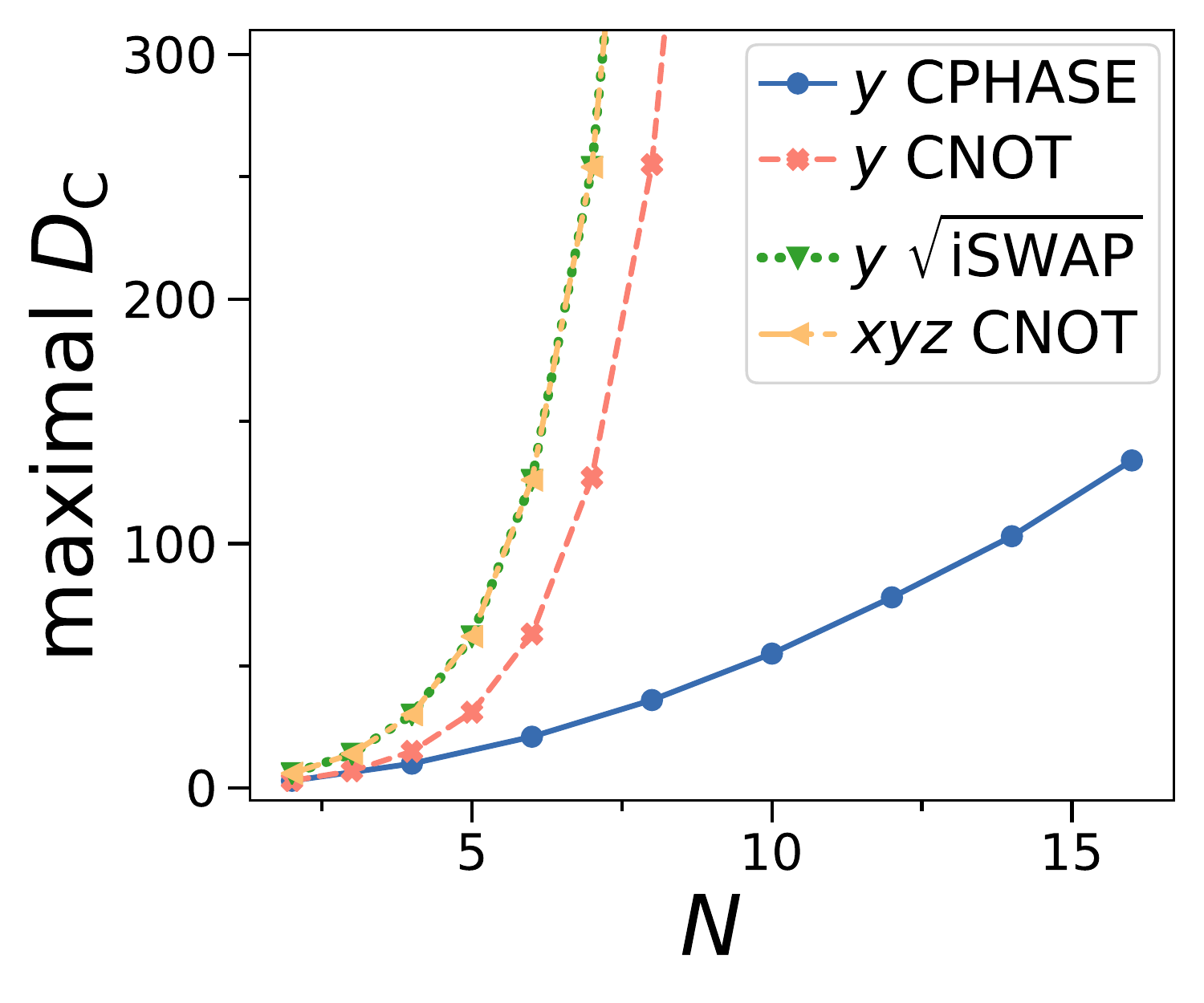}
	\caption{Capacity of PQCs with $y$ rotations and different entangling gates.  The entangling layer is arranged as a nearest-neighbor one-dimensional chain. Three of the PQCs have $y$ rotations, and as reference we show a PQC with randomized $x$, $y$ or $z$ rotations and CNOT gates.
	\idg{a} Variance of the gradient $\text{var}(\partial_k E)$ in respect to the Hamiltonian $H=\sigma_1^z\sigma_2^z$.
	\idg{b} Maximal parameter dimension $D_\text{C}$ of the PQCs as function of number of qubits $N$. For $y$ CPHASE we find an approximate powerlaw $D_\text{C}\propto N^2$.}
	\label{yRotation}
\end{figure}

\begin{figure}[htbp]
	\centering
	\subfigimg[width=0.24\textwidth]{a}{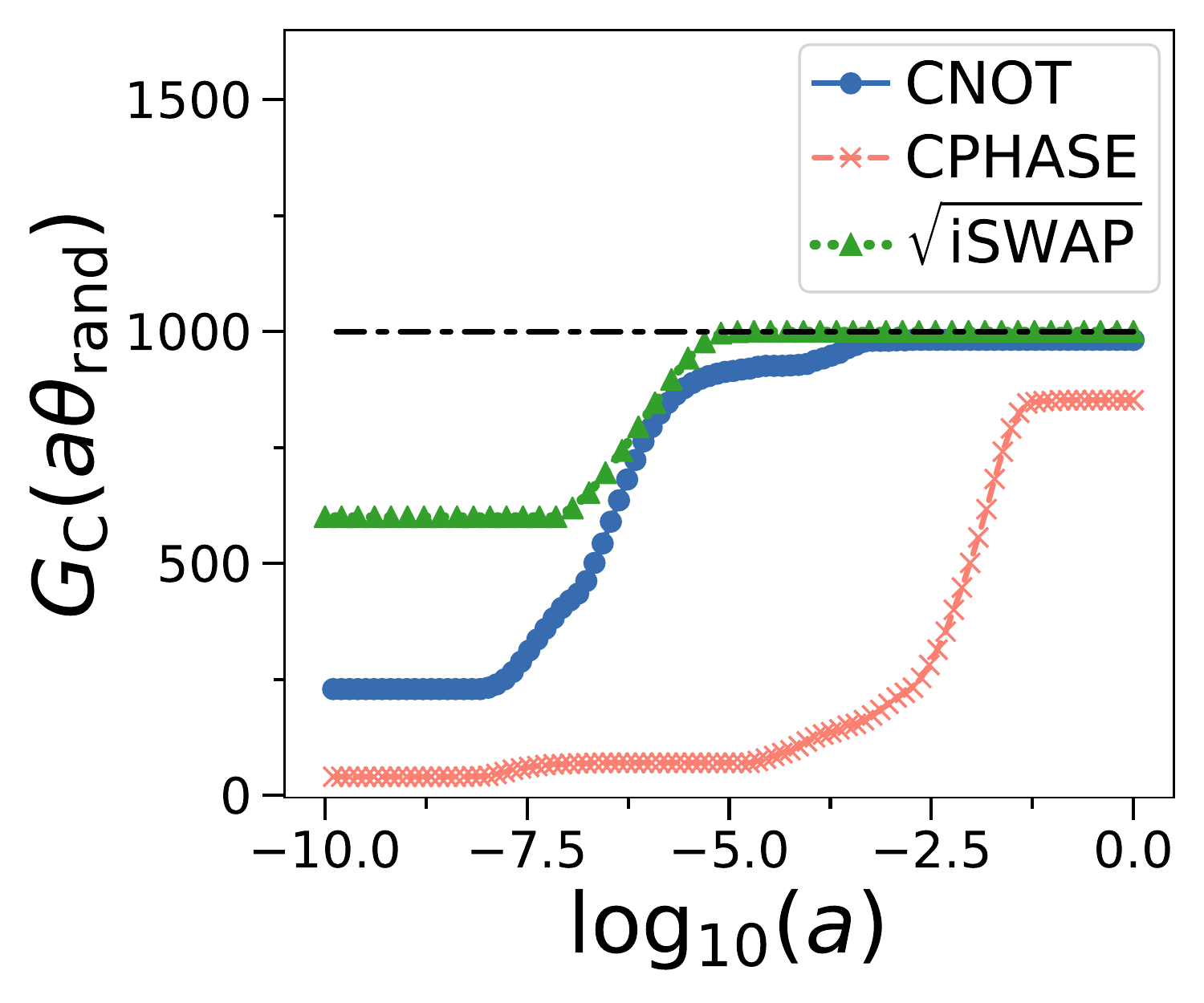}\hfill
	\subfigimg[width=0.24\textwidth]{b}{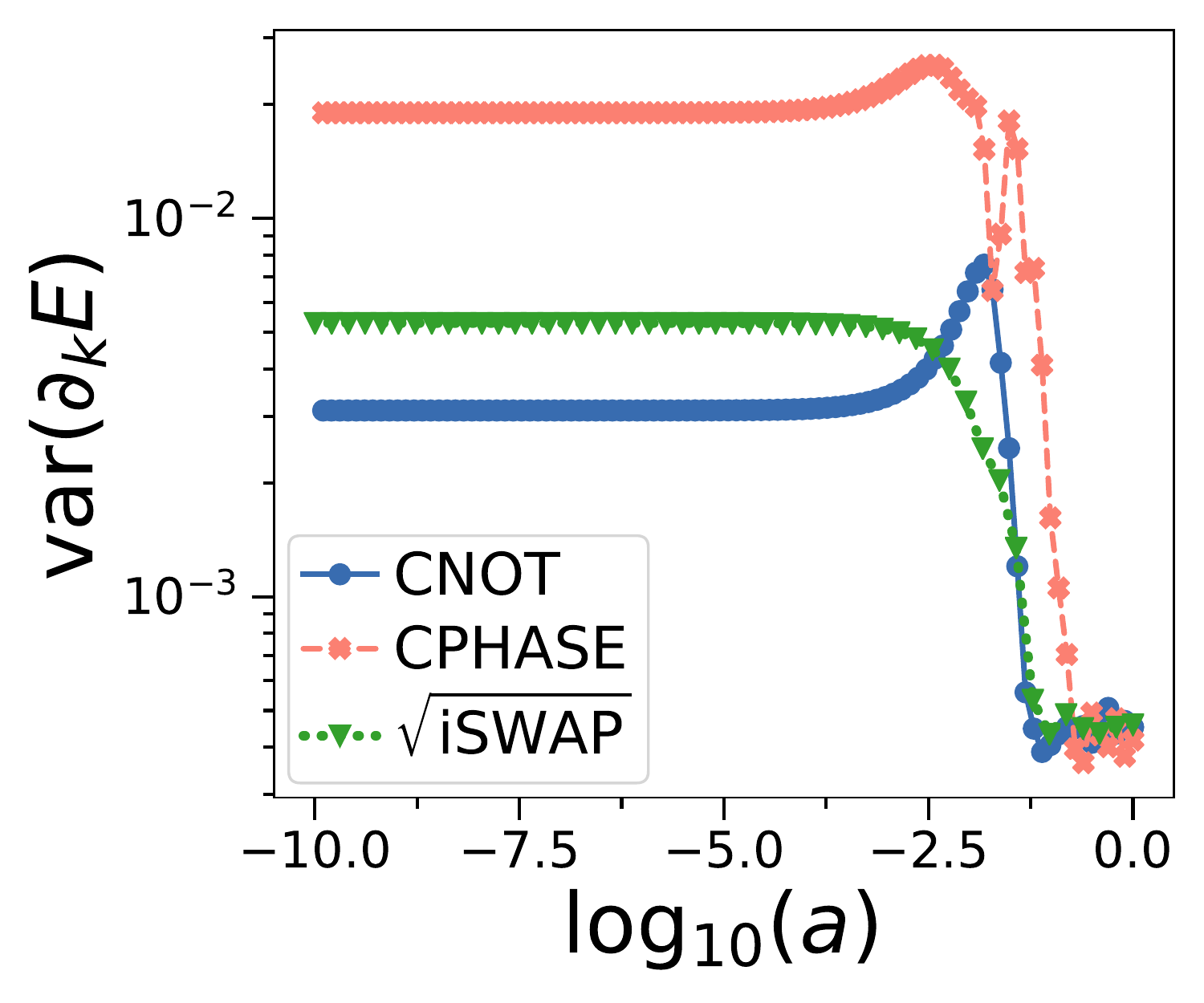}
	\caption{Tuning the PQC parameters $\theta=a\theta_\text{random}$, where $a=(0,1]$ and $\theta_\text{random}\in[0,2\pi)$ for circuits composed of random $x$, $y$ and $z$ rotations and entangling gates arranged in a  chain configurations. 
	\idg{a} The effective quantum dimension $G_\text{C}$ as function of $\log_{10}(a)$. Black dashed-dotted line is number of parameters $M$.
	\idg{b} Variance of the gradient $\text{var}(\partial_k E)$ in respect to Hamiltonian $H=\sigma_1^z\sigma_2^z$.
	All plots show number of layers $p=100$ and $N=10$ qubits.}
	\label{Evolve}
\end{figure}

In Fig.\ref{Evolve} we show how $G_\text{C}$ and the variance of the gradient \revB{for $H=\sigma_1^z\sigma_2^z$} changes when tuning the parameters of a PQC defined as $U(a\theta_\text{random})\ket{0}$, $\theta_\text{random}\in[0,2\pi)$, $a\in[0,1]$. When adjusting $a=0$ to $a=1$, this corresponds to changing the PQC from parameters all zero to a PQC with random parameters. \revB{As example, we show} a PQC consisting of layered randomly chosen single qubit rotations around $x$,$y$,$z$ axis and entangling gates arranged in a chain. 
In Fig.\ref{Evolve}a, we show $G_\text{C}$ for different types of entangling gates. $G_\text{C}$ increases with $a$, reaching the parameter dimension $D_\text{C}$ for $a=1$. CNOT and $\sqrt{\text{iSWAP}}$ increase faster with $a$ compared to the PQC with CPHASE gates.
\revB{In Fig.\ref{Evolve}b,} the variance of the gradient decreases sharply once a particular $a$ is reached. 
Note that there is a specific range of parameters $\text{log}_{10}(a)\approx-2.5$ where the PQCs have nearly maximal $G_\text{C}$ and the variance of gradients remains large. 

\begin{figure}[htbp]
	\centering
	\subfigimg[width=0.24\textwidth]{}{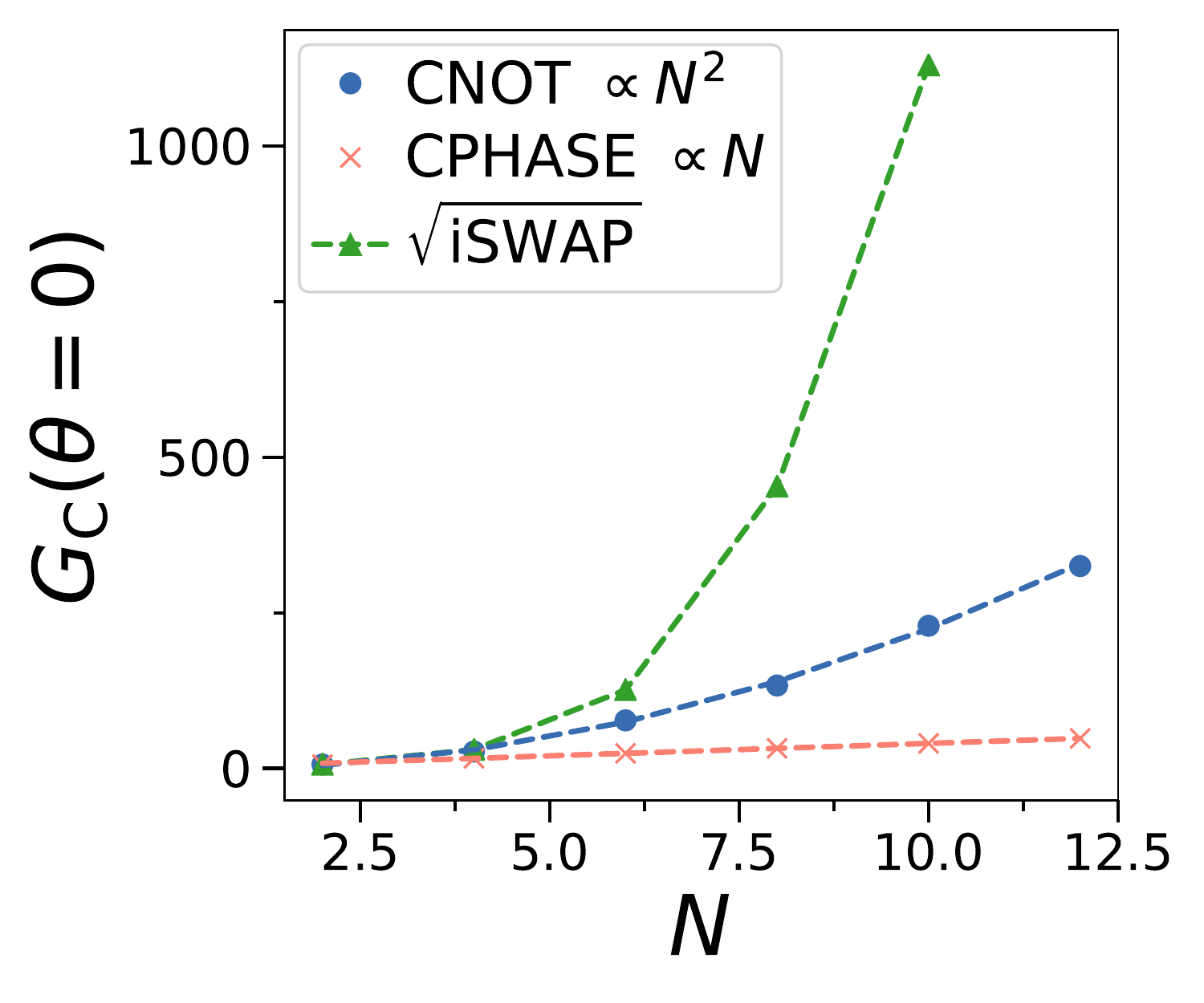}
	\caption{Effective quantum dimension $G_\text{C}(\theta=0)$ plotted against number of qubits $N$ for a circuit consisting of randomly chosen parametrized rotations around $x$, $y$ or $z$ axis with parameters $\theta=0$, and two-qubit entangling gates arranged in a nearest-neighbor chain. We compare CNOT, CPHASE and $\sqrt{\text{iSWAP}}$ entangling gates.  From numerical results, we find $G_\text{C}$ scales quadratically for CNOT gates, linearly for CPHASE gates and higher order polynomial or even exponential scaling for $\sqrt{\text{iSWAP}}$. Number of layers $p$ is chosen such that $G_\text{C}(\theta=0)$ is maximized. }
	\label{ScaleParam0}
\end{figure}

In Fig.\ref{ScaleParam0} we show the scaling of $G_\text{C}(\theta=0,N)$ with number of qubits $N$ for a PQC with entangling gates in a chain arrangement initialized with $\theta=0$, corresponding to the point $a=0$ in Fig.\ref{Evolve}. Numerically, we find linear scaling of $G_\text{C}(\theta=0,N)$ for CPHASE entangling gates, quadratic scaling for CNOT gates and higher order polynomial or even exponential scaling for $\sqrt{\text{iSWAP}}$ gates.

As an application, we propose Algorithm~\ref{alg:algo1} to remove redundant parameters from a PQC $C$. \revB{The algorithm calculates the eigenvectors of the QFI with eigenvalue zero. Parameters which have a non-zero amplitude in the eigenvectors can potentially be removed from the PQC without changing its expressive power. The algorithm removes one redundant gate and removes the corresponding entry in the QFI, then re-calculates the eigenvectors of the QFI. These steps are repeated until no redundant gates are left. }  The resulting pruned PQC $C_\text{pruned}$ has as many parameters as the parameter dimension $D_C$ of the original PQC. \revB{We demonstrate our algorithm on the CPHASE-CHAIN PQC in Appendix~\ref{sec:prune} and find a substantial reduction of parameters without affecting $D_C$.}

\begin{algorithm}[h]
 \SetAlgoLined
 \LinesNumbered
  \SetKwInOut{Input}{Input}
  \SetKwInOut{Output}{Output}
   \Input{\revA{PQC $C$,  QFI $\mathcal{F}^C(\theta_\text{random})$, number of parameters $N_\text{param}^C$, $D_{C}<N_\text{param}^C$,  empty set $\mathcal{K}=\{\}$}}
    \Output{\revA{Pruned PQC $C_\text{pruned}$ with $N_\text{parameters}^{C_\text{pruned}} \approx D_{C}$}}
 \SetKwRepeat{Do}{do}{while}
    \Do{\revA{$\mathcal{F}^C(\theta_\text{random})$ has non-zero eigenvalues}}{
    \revA{Get eigenvalues $\lambda^{(i)}$ of $\mathcal{F}^C(\theta_\text{random})$ sorted in ascending order, eigenvectors $\alpha^{(i)}$ and rank $r$
    
    Calculate $\boldsymbol{\beta}_j=\sum_{i=1}^{N_\text{param}-r} \vert\alpha_j^{(i)}\vert^2$ where $\alpha^{(i)}$ denotes the $i$-th eigenvector with corresponding eigenvalue $\lambda^{(i)}=0$ 
    
    Pick largest index $k$ such that $\boldsymbol{\beta}_k\neq0$
    
    Update $\mathcal{F}^C(\theta_\text{random})$ by removing row $k$ and column $k$ 
    
    Add $k$ to set $\mathcal{K}$}}

    \revA{Removing parameters corresponding to set $\mathcal{K}$ from $C$ gives pruned PQC $C_\text{pruned}$}
    
 \caption{\revA{Prune PQC of redundant parameters}}
 \label{alg:algo1}
\end{algorithm}

\section{Discussion}\label{sec:discussion}
We investigated the capacity and trainability of hardware efficient PQCs using the quantum geometric structure of the parameter space. 
We introduced the notion of parameter dimension $D_\text{C}$ and effective quantum dimension $G_\text{C}$ which are global and local measures respectively of the space of quantum states that can be accessed by the PQC. Both can be derived from the QFI.
We applied these concepts on PQCs composed of layers of single-qubit rotations and different types of entangling gates arranged in various geometries.
For comparable circuit depth $p$, we find strong numerical evidence that PQCs constructed from CNOT or $\sqrt{\text{iSWAP}}$ gates have lower variance of the gradient, and thus higher expressibility compared to PQCs with CPHASE gates. \revB{While two-qubit gates such as CNOT and CPHASE gate can be expressed as each other by applying specific single-qubit rotations, the PQCs we use only have a limited amount of single qubit rotations and thus the choice and arrangement of two-qubit gates strongly affects the expressibility of the PQC.}
\revB{Without loosing generality, we study the properties of the variance of the gradient using a two-qubit Hamiltonian $H=\sigma_1^z\sigma_2^z$, where we take the variance over an ensemble of randomized PQCs. The variance of the gradient of a generic Hamiltonian $H=\sum_i b_i H_i$ that consists of a polynomial number of Pauli operators $H_i$ shows the same exponential decay as the two-qubit Hamiltonian~\cite{mcclean2018barren,cerezo2021cost}, which we demonstrate for a many-body Hamiltonian in the Appendix~\ref{sec:ising_decay}.}

For a specific type of PQC composed of $y$ rotations and CPHASE gates, $D_\text{C}$ scales only quadratically with number of qubits, which may imply that this PQC can be efficiently simulated on classical computers. 
We find that the redundancy of parameters varies strongly depending on the configuration of the PQC as well as the type of gates.

The effective quantum dimension $G_\text{C}$ reveals the expressive power of a PQC by local variations around a specific parameter set. We find that depending on the entangling gates, $G_\text{C}$ shows widely different scaling with number of qubits, with the largest value found for $\sqrt{\text{iSWAP}}$ gates. 
While we only studied the case $\theta=0$, PQCs with correlated parameters could feature similar behavior~\cite{volkoff2020large}. 
Tuning the parameters of a PQC from zero to a random set of parameters yields a crossover from large gradients and small $G_\text{C}$ to vanishing gradients and large $G_\text{C}$. For the PQCs investigated, we can find a range of parameters that combines large gradients with a nearly maximal $G_\text{C}$, which could be an optimal starting point for gradient based optimisation.
Trade-offs between the expressibility of a circuit and the magnitude of its gradients are a key challenge in finding good initialization strategies~\cite{holmes2021connecting}.

When increasing the number of layers $p$ to a value $p_\text{c}$, a transition occurs in the QFI when $D_\text{C}$ reaches its maximal possible value. The transition is characterized by a disappearance of small eigenvalues of the QFI and a peak in the variance of the logarithm of eigenvalues. 
This peak may be related to a transition in the optimization landscape of control theory. When the system becomes overparameterized with more parameters than degrees of freedom, the optimization landscape changes from being spin-glass like with many near-degenerate minima to one with many degenerate global minima~\cite{bukov2018reinforcement,rabitz2004quantum}.
\revB{This peak in the QFI could be used to identify the transition.}
The overparameterized regime may be useful for mitigating the effect of noise~\cite{fontana2020evaluating,fontana2020optimizing}.
For deep circuits $p>p_\text{c}$, the transition leads to a decay of the QNG as small eigenvalues are suppressed.
For shallow circuits $p<p_\text{c}$, the QNG can be orders of magnitude larger in value compared to the regular gradient, however our numerical results suggest that both regular gradient and QNG decrease exponentially with number of qubits. 
Thus, the QNG most likely cannot help to solve the barren plateau problem. This contrasts the natural gradient in classical machine learning, which is known to be able to overcome the plateau phenomena that leads to a slow down of optimization~\cite{amari2016information}.

Imaginary-time evolution and variational quantum simulation use a matrix related to the QFI to update the parameters of the PQC~\cite{mcardle2019variational,stokes2020quantum}. The effective quantum dimension $G_\text{C}$ could give major insights on the convergence properties of these algorithms. 
Recent proposals for adaptively generated ans\"atze could benefit from the QFI by taking the geometry of the PQC into account when designing PQCs~\cite{grimsley2019adaptive}.

We demonstrated an algorithm to systematically reduce the number of parameters and depth of PQCs while keeping the parameter dimension constant. This algorithm can be immediately applied to PQCs used in VQAs to reduce the number of parameters $M$ without sacrificing expressive power. Commonly used PQCs often contain more parameters than necessary. \revC{Removing them reduces the computational effort for calculating the gradient as well as the QFI necessary for the QNG, which has been shown to be highly beneficial for training~\cite{wierichs2020avoiding,van2020measurement}. When compared to ordinary gradient descent, the sampling overhead of training using the QFI and QNG is constant asymptotically for both an increasing number of iterations and number of qubits, as has been proven recently~\cite{van2020measurement}. Furthermore, training with the QNG has a reduced total cost since it approaches the optimum faster~\cite{van2020measurement}. Thus, NISQ algorithms that use the QFI to update their parameters accomplish faster training than ordinary gradient descent.  Our algorithm reduces the cost of calculating QFI in each iteration of the training by truncating the size of the QFI. As the QFI is a matrix, removing a single parameter already reduces the number of elements to measure by $M$.} 
Further, with our approach one can lower the number of parameterized gates needed to run the VQA, which is especially important for NISQ era algorithms. 

\revC{The QFI has widespread use in quantum metrology~\cite{liu2019quantum} and quantum computing~\cite{stokes2020quantum,yao2020adaptive,meyer2021fisher}. 
To facilitate its application, various methods to calculate the QFI on quantum computers have been developed and are continuously improved~\cite{meyer2021fisher,wierichs2021general}, which we review in Appendix~\ref{sec:meas_qfi}.
The most commonly applied methods are the shift-rule~\cite{mari2021estimating,wierichs2021general}, the Hadamard test~\cite{li2017efficient,yuan2019theory,yao2020adaptive} and direct measurement methods~\cite{mitarai2019methodology}. For these approaches, the number of circuits to measure scales as the square $\left(M^2\right)$ of the number of parameters $\left(M\right)$.
Various approximations for the QFI have been proposed~\cite{stokes2020quantum,gacon2021simultaneous,cerezo2021sub,rath2021quantum}. 
Improved methods for numerical simulation of the QFI are being developed as well~\cite{jones2020efficient}. We provide code that can simulate the QFI for 26 qubits on a desktop computer~\cite{haug2021quantumgeometry}. 
We note that calculations relying on a reduced number of qubits or layers can help to design better PQCs. Most commonly used PQCs are constructed according to specific rules in a layer-wise fashion. By evaluating the effective dimension within smaller PQCs, one can identify rules and patterns for constructing PQCs with few redundant parameters. Then, one can extrapolate these rules to PQCs with many qubits and layers.
}

During the training of a PQC, the eigenvalue spectrum of the QFI can gain specific features, as has been shown for restricted Boltzmann machines~\cite{park2020geometry}. We show that the PQCs have a characteristic eigenvalue spectra depending on the type of gates and their arrangement (see Appendix~\ref{sec:further_eigenvalues}). 
The eigenvalues hold important information about the trainability and generalization of a model. For example, a model that generalizes well is known to have a low effective dimension in classical machine learning~\cite{maddox2020rethinking}. It would be interesting to study in what way these statements translate to quantum machine learning. The eigenvalues of the Hessian could be applied as well~\cite{huembeli2021characterizing}.
Further, connections to complementary measures of capacity based on classical Fisher information~\cite{abbas2020power} and memory capacity~\cite{wright2020capacity} respectively could be explored.

\revB{While we studied hardware efficient PQCs, some of our results can be carried over to other types of PQCs. The transition in the QFI spectrum we observed could be used to characterize when a PQC is overparameterized. Further, $G_C(\theta)$ can be used to determine the amount of quantum states that can be reached by varying the parameters of PQCs. }
It would be straightforward to extend our concepts to evaluate the capacity and trainability of noisy PQCs~\cite{koczor2019quantum}, convolutional PQCs~\cite{cong2019quantum}, optimal control~\cite{magann2020pulses}, quantum metrology~\cite{meyer2020variational} and programmable analog quantum simulators~\cite{bastidas2020fully}.

Python \revB{and Julia} code for the numerical calculations performed in this work are available at~\cite{haug2021quantumgeometry}.

\medskip
{\noindent {\em Acknowledgements---}} This work is supported by a Samsung GRC project and the UK Hub in Quantum Computing and Simulation, part of the UK National Quantum Technologies Programme with funding from UKRI EPSRC grant EP/T001062/1. We are grateful to the
National Research Foundation and the Ministry of Education, Singapore for financial support. 
\bibliographystyle{apsrev4-1}
\bibliography{GeometricTensor}

\appendix 

\section{Variational quantum eigensolver}
The core idea of Variational quantum eigensolver (VQE) is to find the ground state of a Hamiltonian $H$ by minimizing the parameters $\theta$ of a PQC in regards to an objective function that represents the energy of a given Hamiltonian $E(\theta)=\bra{0}U^\dagger(\theta)HU(\theta)\ket{0}$~\cite{peruzzo2014variational}. The minimisation is performed with a classical optimisation algorithm, whereas the energy is measured on a quantum device. According to the Ritz variational principle, the objective function is lower bounded by the ground state energy of $H$, i.e. $E(\theta)\ge E_\text{g}$, where $E_\text{g}$ is the true ground state of $H$.

\section{Quantum Fisher information metric}\label{sec:Fisherinfo}
For VQE, the objective function is updated in hybrid classical-quantum algorithm in an iterative manner. At step $n$ of the procedure, the objective function is evaluated on the quantum computer for a given $\theta_n$. Based on the result, a classical computer selects the next choice $\theta_{n+1}$ such that it (hopefully) decreases the objective function. A common scheme to update parameters is ordinary gradient descent
\begin{equation}
\label{standard gradient}
      \theta_{n+1} = \theta_n - \eta \frac{\partial E(\theta)}{\partial \theta}\, ,
\end{equation}
where $\eta$ is a small coefficient and $\partial E(\theta)/\partial \theta$ is the gradient of the objective function. 

The above update rule assumes that the parameter space for $\theta$ is a flat Euclidian space. However, in general this is not the case, as the underlying PQC and cost function do not have such simple forms.
Recent studies have proposed the quantum natural gradient (QNG), inspired from the natural gradient in classical machine learning~\cite{amari1998natural}, to minimize the objective function~\cite{yamamoto2019natural,stokes2020quantum}.
The main idea is to use information about how fast the quantum state changes when adjusting the parameter $\theta$ in a particular direction. 
Optimisation with the natural gradient updates the parameters according to
\begin{equation}
\label{natural gradient}
      \theta_{k+1} = \theta_k - \eta_k \mathcal{F}^{-1}(\theta)\frac{\partial E(\theta)}{\partial \theta}, 
\end{equation}
where $\mathcal{F}(\theta)$ is the Fubini-Study metric tensor or quantum Fisher information metric (QFI)
\begin{equation}
\mathcal{F}_{ij}=\text{Re}(\braket{\partial_i \psi}{\partial_j \psi}-\braket{\partial_i \psi}{\psi}\braket{\psi}{\partial_j \psi})\label{eq:Fisher}\,,
\end{equation}
where $\ket{\partial_i \psi}=\frac{\partial}{\partial \theta_i}\ket{\psi(\theta)}$ denotes the partial derivative of $\ket{\psi(\theta)}$. 
One can relate $\mathcal{F}(\theta)$ to the distance in the space of pure 
quantum states, which is the Fubini-Study distance given by 
\begin{equation}\label{eq:fdistSupp}
    {\rm Dist}_Q\Big(\ket{\psi(\theta)}, ~ \ket{\psi(\theta+\text{d}\theta)} \Big)^2
      = \sum_{i,j}\mathcal{F}_{ij}(\theta)\text{d}\theta_i \text{d}\theta_j\,, 
\end{equation}
where ${\rm Dist}_Q(x,y)=\vert\braket{x}{y}\vert^2$.

\section{Further data on the PQCs}\label{sec:further_data}
\begin{figure}[htbp]
	\centering
	\subfigimg[width=0.24\textwidth]{a}{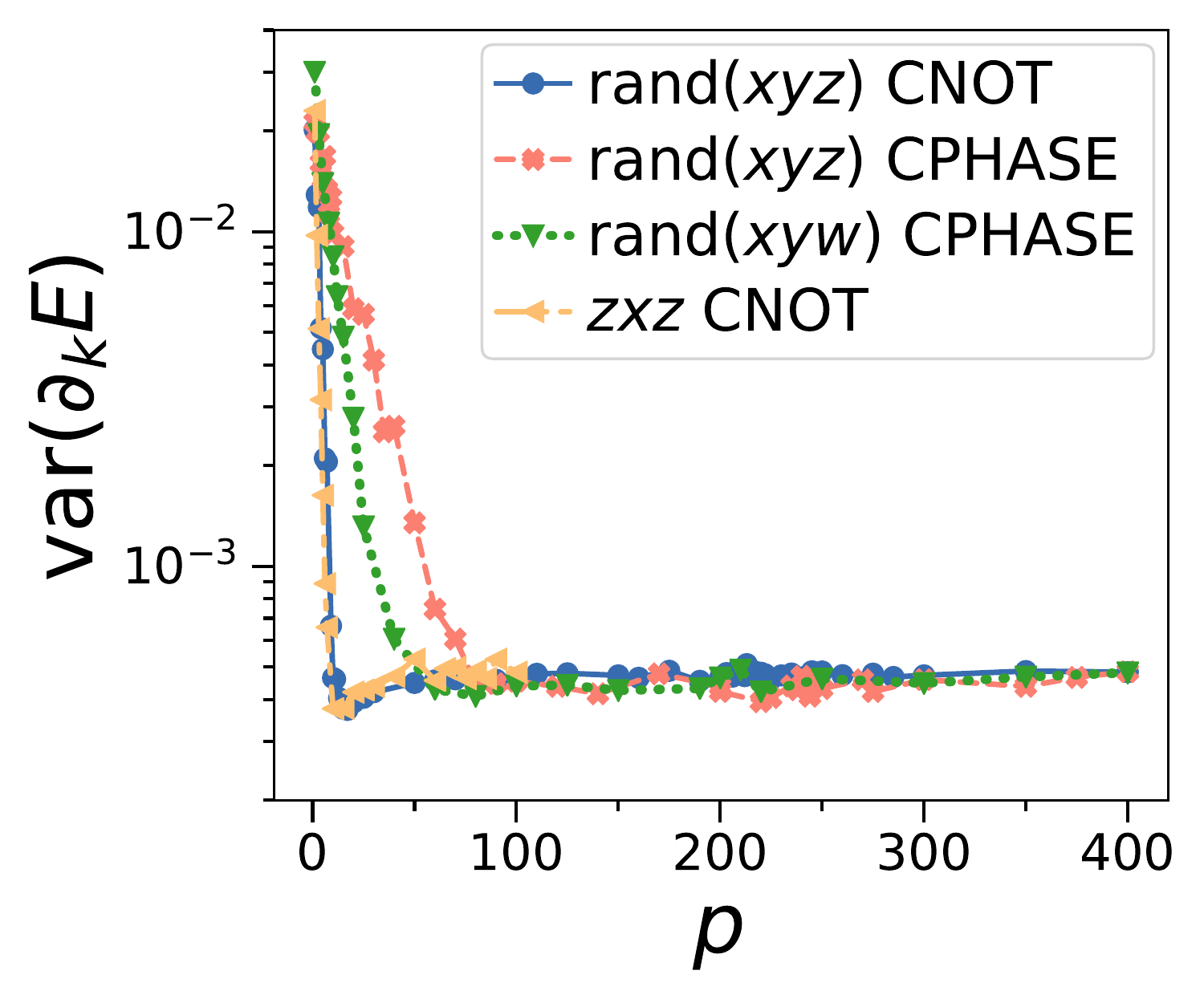}\hfill
	\subfigimg[width=0.24\textwidth]{b}{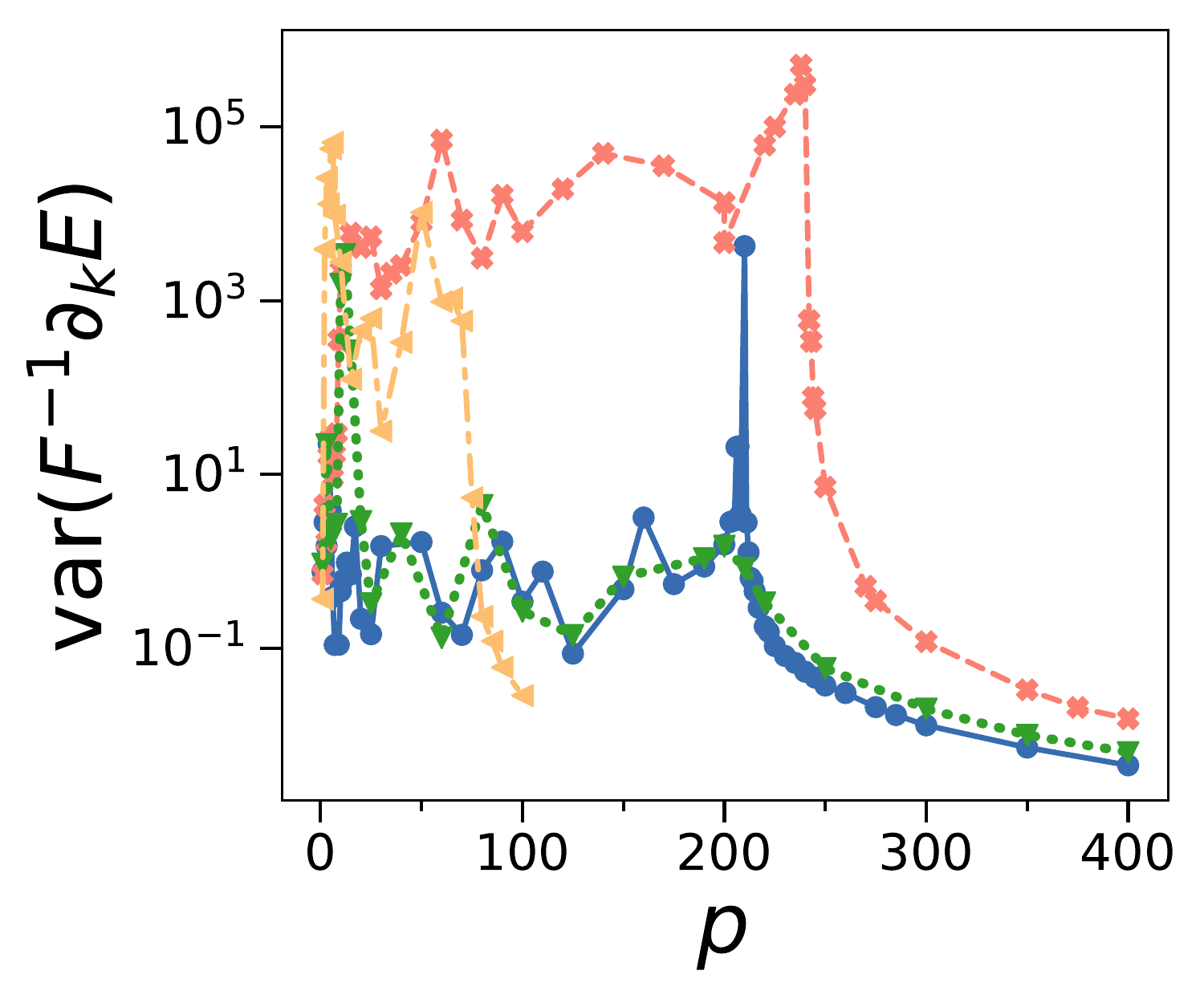}\\
	\subfigimg[width=0.24\textwidth]{c}{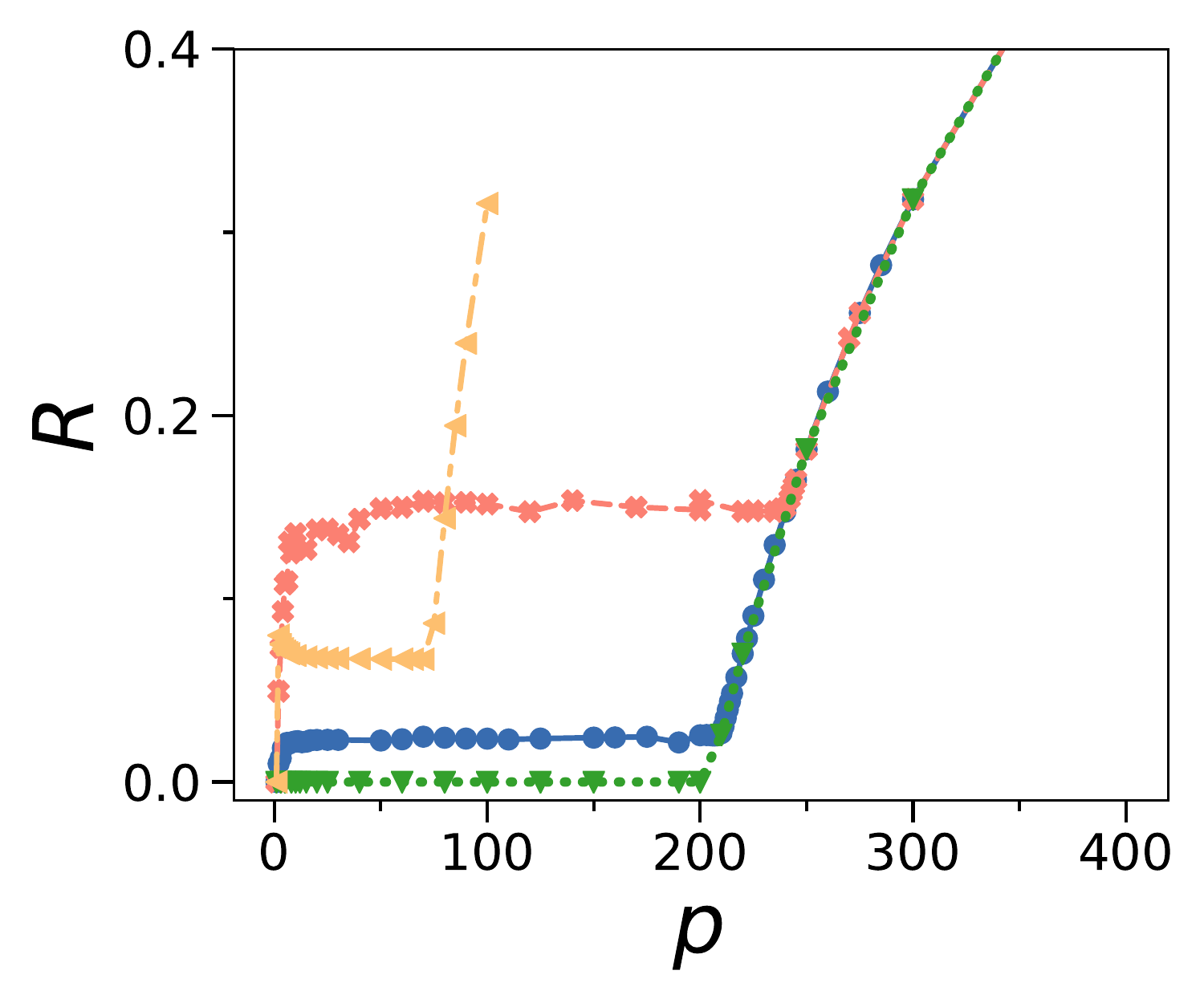}\hfill
	\subfigimg[width=0.24\textwidth]{d}{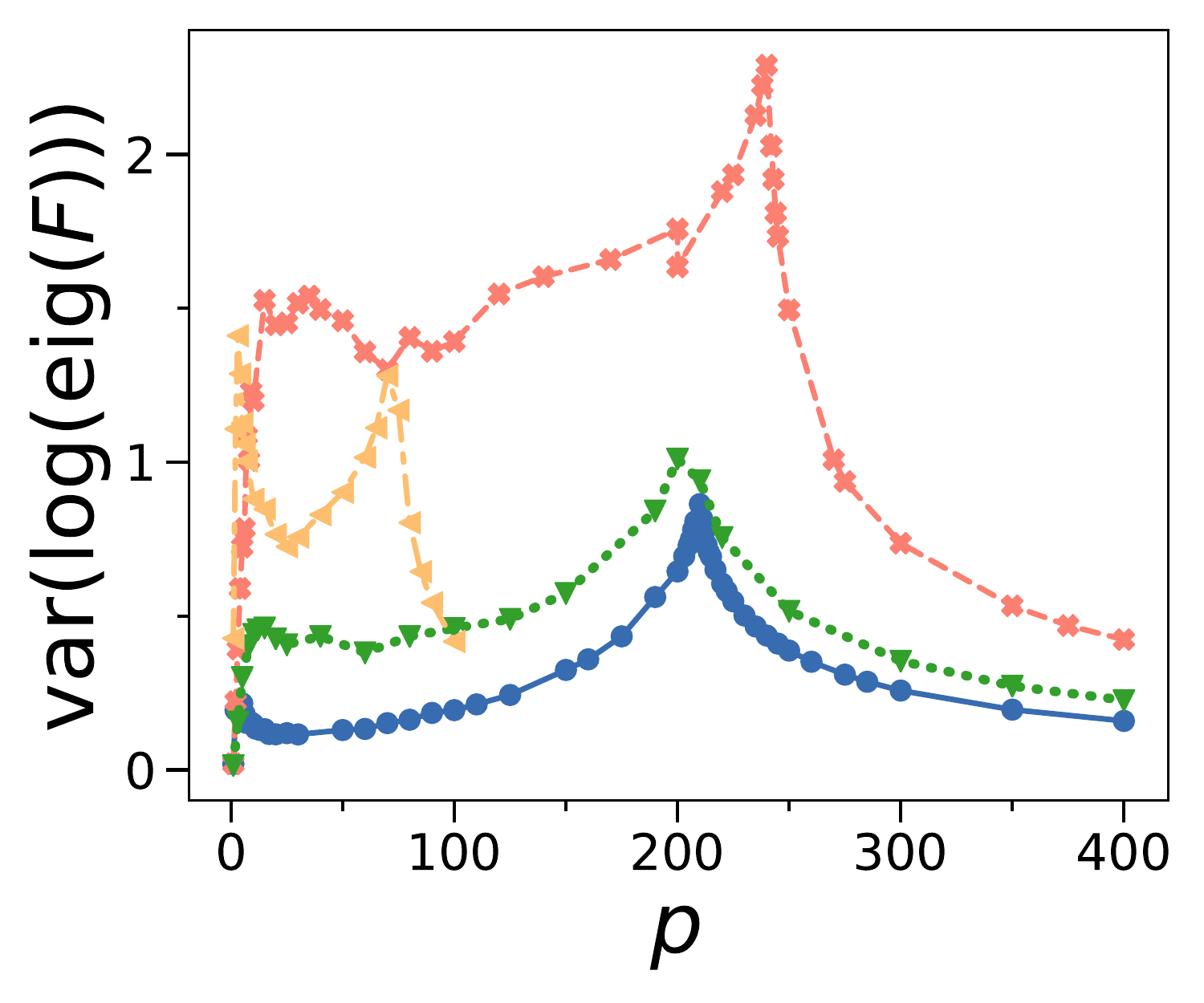}
	\caption{Properties of further PQCs plotted against layers $p$ for $N=10$ qubits. PQCs have nearest-neighbor chain entangling layers. We define the type of PQCs in the legend: rand($xyz$) denotes randomized single-qubit rotations around $\{x,y,z\}$ axis. rand($xyw$) denotes randomized single-qubit rotations around $\{x,y,(x+y)/\sqrt{2})\}$ axis. $zxz$ denotes that for every layer there are three single-qubit rotations, around $z$, $x$ and $z$ axis. 
	\idg{a} Variance of the gradient $\text{var}(\partial_k E)$  in respect to the Hamiltonian $H=\sigma_1^z\sigma_2^z$. 
	\idg{b} Variance of QNG $\text{var}(\mathcal{F}^{-1}\partial_k E)$.
	\idg{c} Redundancy $R$ of parameters of the PQCs.
	\idg{d} Variance of the logarithm of the eigenvalues of the QFI.
	}
	\label{fig:morePQC}
\end{figure}

In Fig.\ref{fig:morePQC}, we show further types of PQCs as defined in the caption. We highlight that the PQC rand($xyw$) CPHASE has lower redundancy compared to rand($xyz$) CPHASE. The reason is that the $z$ rotations, which can commute with the CPHASE layer, are replaced with non-commuting $(x+y)/\sqrt{2}$ rotations. This leads to a faster decrease in the variance of the gradient as well. We also define a common type of PQC $zxz$ CNOT, which has first been introduced in~\cite{kandala2017hardware}. We note that while it has three rotations per qubit and layer, compared to rand($xyz$) CNOT the decay of the variance of the gradient as function of $p$ remains the same in both types of PQC.
Finally, we show further examples of the transition in the QFI, visible both in the peak of the variance of the logarithm of the eigenvalues, and in the decay of the QNG.

\begin{figure}[htbp]
	\centering
	\includegraphics[width=0.28\textwidth]{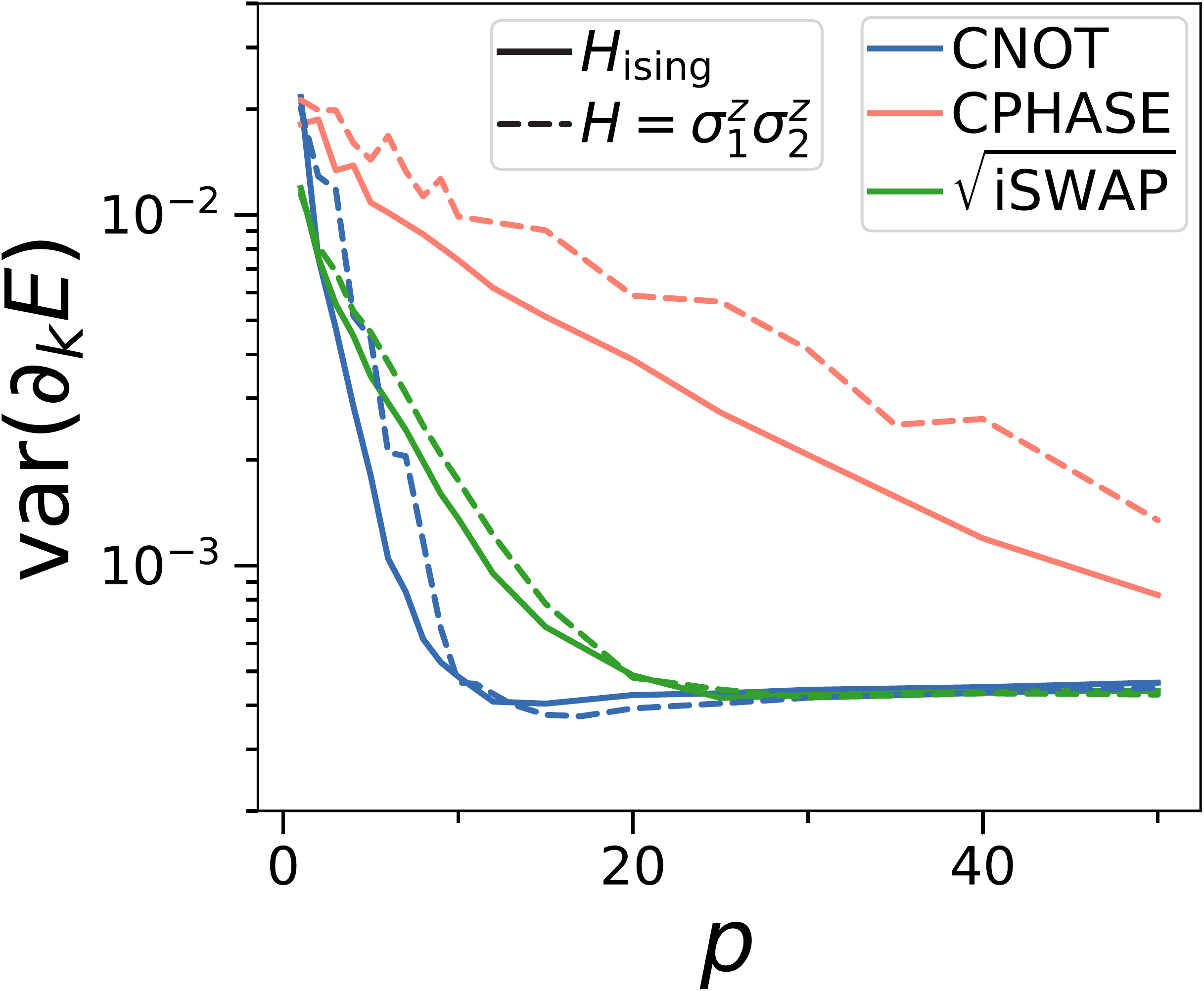}
	\caption{Variance of gradient for $H=\sigma_1^z\sigma_2^z$ (solid curves) and transverse Ising Hamiltonian (dashed curves) for different types of PQC with entangling gates in a CHAIN configuration for varying number of layers $p$. For the transverse Ising Hamiltonian, we divided the variance of the gradient by the number of terms in the Hamiltonian. We use $N=10$ qubits.  }
	\label{compHamiltonian}
\end{figure}

\revB{\section{Variance of gradient of Hamiltonians}\label{sec:ising_decay}
The variance of the gradient shows the same exponential decay due to barren plateaus for any Hamiltonian that consists of a sum of a polynomial number of Pauli operators~\cite{mcclean2018barren}. To demonstrate this, we compare the simple two-qubit Hamiltonian $H=\sigma_1^z\sigma_2^z$ and the transverse Ising model 
\begin{equation}
H_\text{ising}=\sum_{n=1}^N \sigma_n^z\sigma_{n+1}^z+h\sum_{n=1}^N \sigma_n^x
\end{equation}
with $h=1$. The variance of the gradient in respect to the Hamiltonian for different PQCs is shown in Fig.\ref{compHamiltonian}.
We find that the variance of the gradient divided by the number of terms in the Hamiltonian has nearly the same value for both the two-qubit Hamiltonian and the transverse Ising Hamiltonian.
As we take the variance over an ensemble of randomized PQCs, it does not matter which Pauli operator we use to calculate the variance.}

\section{Histograms of eigenvalues}\label{sec:further_eigenvalues}
\begin{figure*}[htbp]
	\centering
	\subfigimg[width=0.24\textwidth]{a}{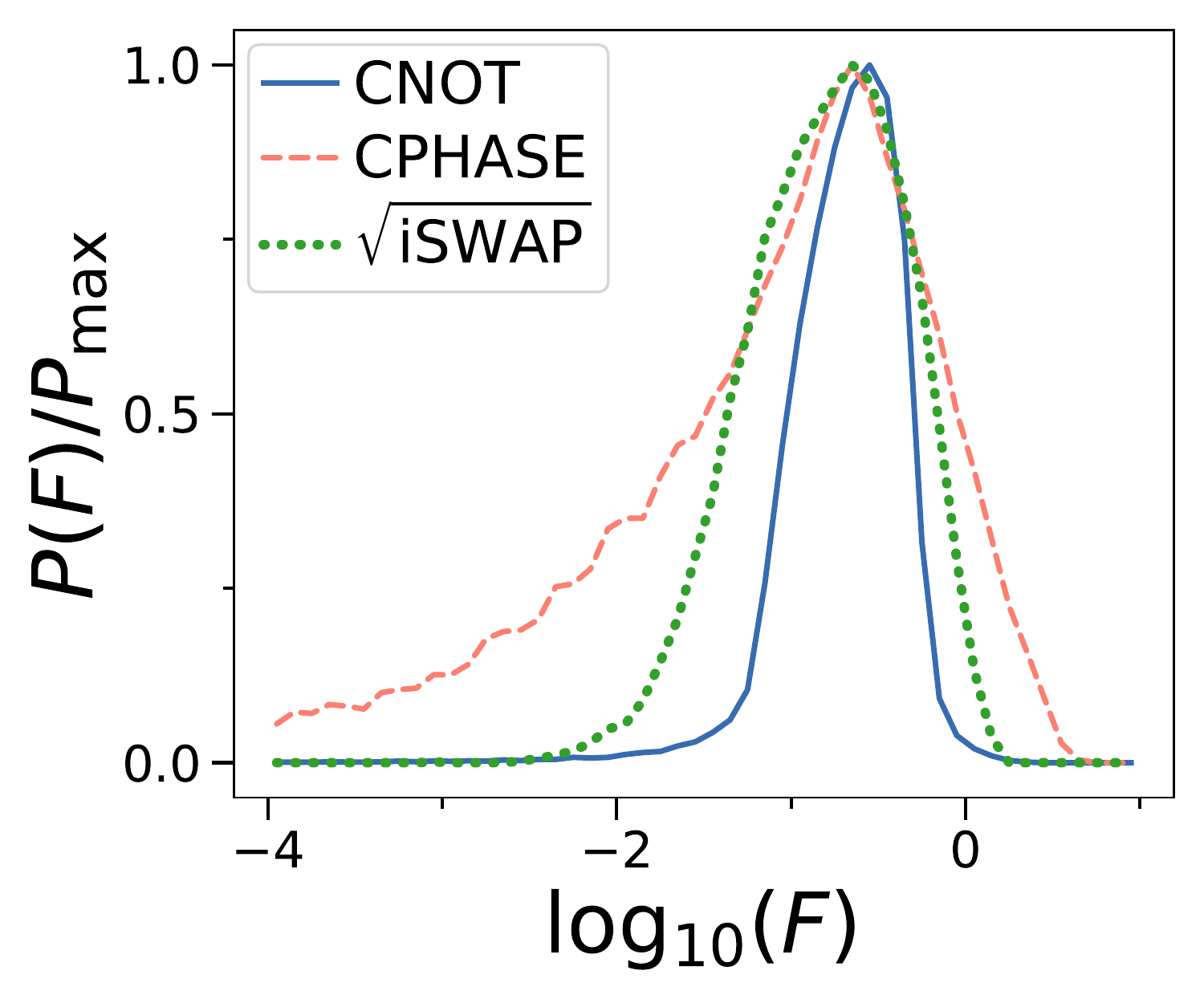}
	\subfigimg[width=0.24\textwidth]{b}{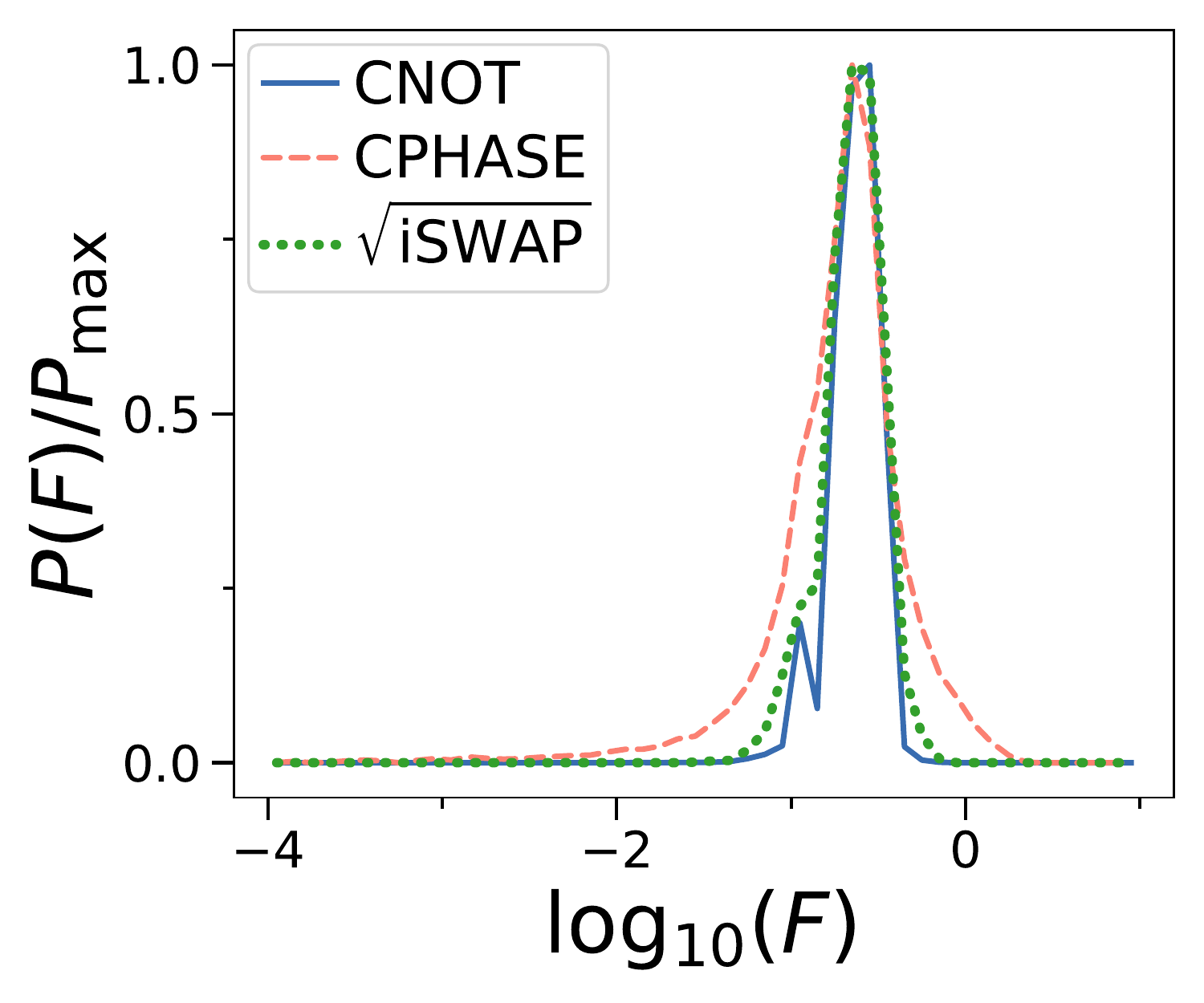}
	\subfigimg[width=0.24\textwidth]{c}{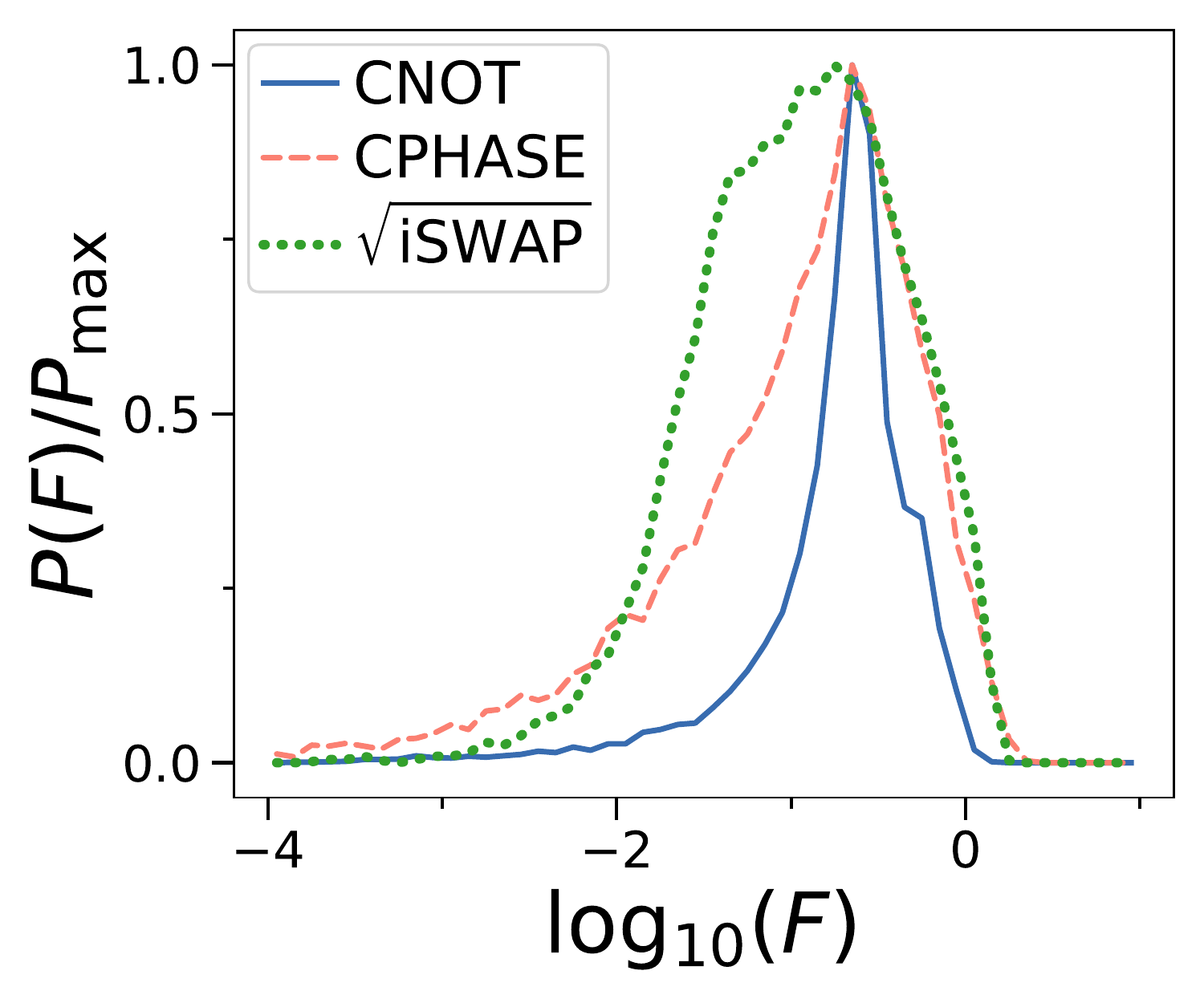}
	\caption{Distribution of eigenvalues of QFI for PQCs shown in Fig.4 in main text. \idg{a} nearest-neighbor chain arrangement of entangling gates \idg{b} all-to-all connectivity \idg{c} alternating nearest-neighbor. All graphs for $N=10$ qubits and number of layers $p=50$.
	}
	\label{fig:moreEigv}
\end{figure*}

In Fig.\ref{fig:moreEigv} we show the distribution of eigenvalues for the PQCs of Fig.4 in the main text. We find that a characteristic spectrum for the different PQC types. Note that CPHASE appears to have more pronounced tails in all cases.

\section{Pruning PQCs of redundant parameters}\label{sec:prune}
\begin{figure}[htbp]
	\centering
	\includegraphics[width=0.49\textwidth]{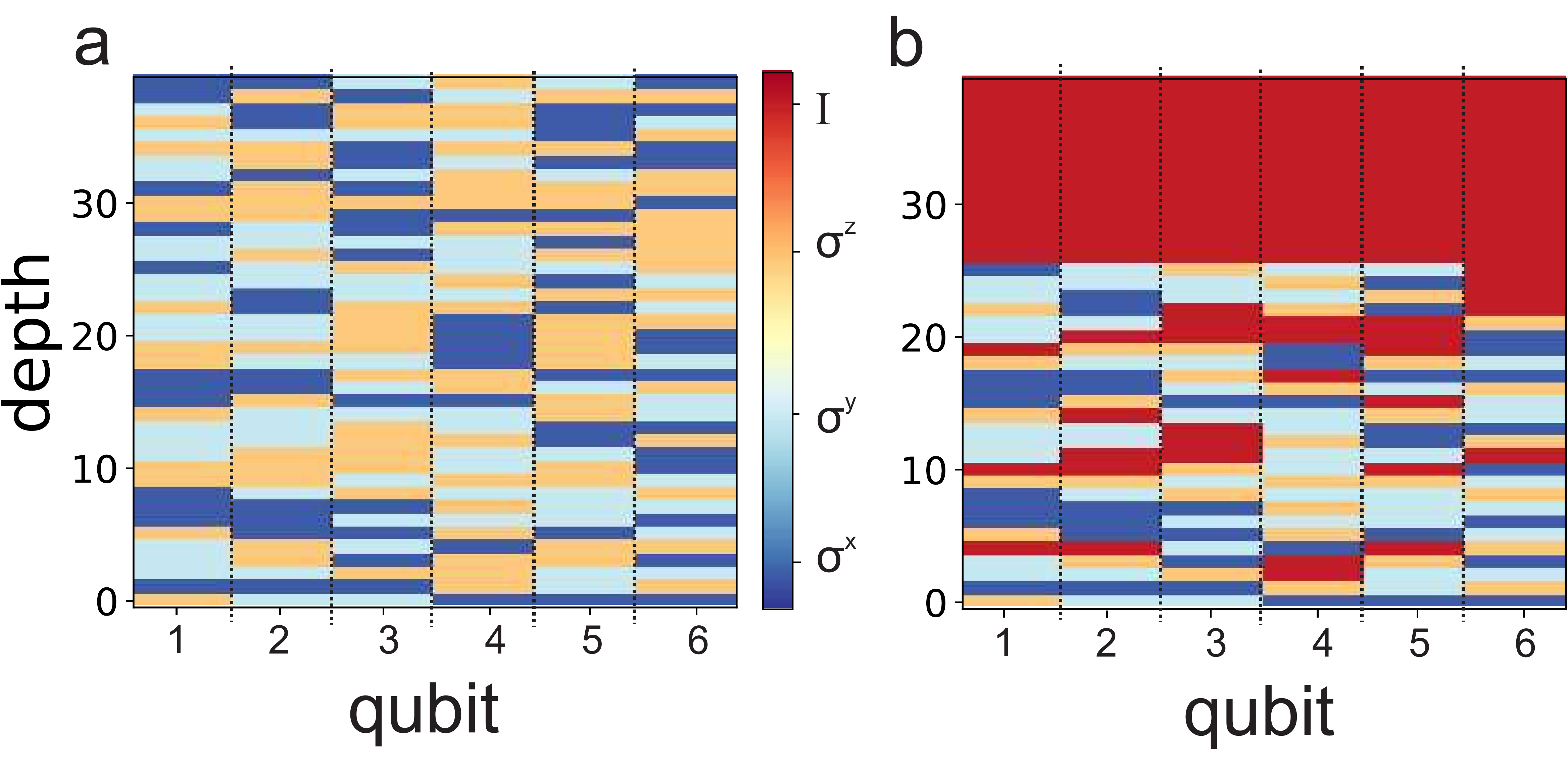}
	\caption{\revA{Pruning of layered PQC composed of random parameterized rotations with Pauli operators $\{\sigma^x,\sigma^y,\sigma^z\}$ and CPHASE gates arranged in a nearest-neighbor chain. PQC has $p=40$ layers and $N=6$ qubits. The colour scheme indicates the type of parameterized rotation at a specific qubit and layer. The identity operation $I$ is to show that no operation (identity) is applied at that particular qubit and depth, which results in pruning the redundant rotations.
	\idg{a} We show the initial PQC with $N_\text{param}=240$ parameters and parameter dimension $D_C=126$. \idg{b} Algorithm 1 removes redundant parameterized rotations and replaces them with identity operator $I$. We show the pruned PQC with $N_\text{param}=126$ and same parameter dimension $D_C=126$, reducing the PQC by 14 circuit layers and 114 parameters.}  }
	\label{PrunePQC}
\end{figure}

\revA{We apply the Algorithm 1 of the main text to prune a PQC of redundant parameters, i.e. parameters which can be removed without changing the parameter dimension $D_C$ and thus the expressiveness of the circuit. 
In Fig.\ref{PrunePQC}a, we show the initial PQC, which is the CPHASE CHAIN PQC. In Fig.\ref{PrunePQC}b, we apply Algorithm 1 of the main text to remove redundant parameters, and reduce the number of unitaries within the circuit substantially. The parameter dimension $D_C=D_{C_\text{pruned}}=126$ remains constant before and after pruning.}

\section{Measuring the quantum Fisher information metric}\label{sec:meas_qfi}
\revC{The QFI has found widespread use in quantum metrology~\cite{liu2019quantum} and quantum computing~\cite{stokes2020quantum,li2017efficient,meyer2021fisher}. 
As such, various methods to calculate the QFI have been proposed and are continuously improved and developed.
We now proceed to review methods for calculating the ${M\times M}$ dimensional positive-semidefinite QFI $\mathcal{F}(\boldsymbol{\theta})$ or Fubini-Study metric
\begin{equation}\label{eq:QFI_sup}
\mathcal{F}_{ij}(\theta)=\text{Re}(\braket{\partial_i \psi}{\partial_j \psi}-\braket{\partial_i \psi}{\psi}\braket{\psi}{\partial_j \psi})\,.
\end{equation}

\emph{Shift-rule---}
The QFI for pure states can be reformulated as the second order derivative of the fidelity of a quantum state~\cite{wierichs2021general}
\begin{equation}
\mathcal{F}_{ij}(\boldsymbol{\theta})=-\frac{1}{2}\partial_i\partial_j\abs{\braket{\psi(\boldsymbol{\theta})}{\psi(\boldsymbol{\theta}_0)}}^2\Big|_{\boldsymbol{\theta}=\boldsymbol{\theta}_0}\,.
\end{equation}
A straightforward way to calculate the QFI is thus calculating the Hessian of the fidelity. 
First, we explain how to calculate fidelities using the inversion test.
The fidelity $K=\abs{\braket{\psi_1}{\psi_2}}^2$ of two quantum states $\ket{\psi_1}=U(\boldsymbol{\theta}_1)\ket{0}$ and $\ket{\psi_2}=U(\boldsymbol{\theta}_2)\ket{0}$ is computed by preparing the first state followed by the inverse of the second state
$\ket{\psi_{12}}=U^\dagger(\boldsymbol{\theta}_2)U(\boldsymbol{\theta}_1)\ket{0}$. Then, the fidelity is measured as the probability of sampling the all-zero state $\abs{\braket{\psi_1}{\psi_2}}^2=\abs{\braket{0}{\psi_{12}}}^2$.
Now that we know how to calculate the fidelity, we can now proceed to calculate its gradients as well.
For quantum computers, the shift rule is a practical way to calculate gradients~\cite{mitarai2018quantum}. 
It directly applies to all PQCs where the parameterized rotations are of the form $\exp(-i\theta G)$, where the generator $G$ is a Pauli string. Recently, the shift-rule has been also extended to circuits with general generators $G$~\cite{wierichs2021general,kyriienko2021generalized,izmaylov2021analytic}. 
The shift-rule for the QFI takes the following form~\cite{mari2021estimating,wierichs2021general}
\begin{align*}
\mathcal{F}_{ij}(\boldsymbol{\theta})=-\frac{1}{8}&[\abs{\braket{\psi(\boldsymbol{\theta}}{\psi(\boldsymbol{\theta}+(\boldsymbol{e}_i+\boldsymbol{e}_j)\pi/2}}^2+\\
&-\abs{\braket{\psi(\boldsymbol{\theta}}{\psi(\boldsymbol{\theta}+(\boldsymbol{e}_i-\boldsymbol{e}_j)\pi/2}}^2\\
&-\abs{\braket{\psi(\boldsymbol{\theta}}{\psi(\boldsymbol{\theta}+(-\boldsymbol{e}_i+\boldsymbol{e}_j)\pi/2}}^2\\
&+\abs{\braket{\psi(\boldsymbol{\theta}}{\psi(\boldsymbol{\theta}-(\boldsymbol{e}_i+\boldsymbol{e}_j)\pi/2}}^2]\,,
\end{align*}
where $\boldsymbol{e}_i$ is the basis vector for $i$-th index of parameter $\boldsymbol{\theta}$.
The diagonal elements of the QFI simplify to
\begin{equation}
\mathcal{F}_{ii}(\boldsymbol{\theta})=\frac{1}{4}[1-\abs{\braket{\psi(\boldsymbol{\theta}}{\psi(\boldsymbol{\theta}+\boldsymbol{e}_i\pi}}^2\,.
\end{equation}
To determine the full QFI, we have to measure in total $2M(M-1)+M$ fidelities via the shift-rule.

Approximations of the QFI can be measured even more efficiently on quantum computers~\cite{stokes2020quantum,gacon2021simultaneous}. 
For example, when unitaries within the PQC commute, one can use this to speed up the calculation. This is the case for the diagonal and block-diagonals entries of the QFI, which can be calculated in a time that scales linearly with $M$~\cite{stokes2020quantum,wierichs2021general}.

\emph{Hadamard test---} 
We now review an alternative approach to calculate the QFI.
We assume a general ansatz for the PQC for $N$ qubits and $M$ parameters
\begin{equation}
\ket{\psi(\theta)}=\prod_{l=1}^M W_l R_l(\theta_l)\ket{0}^{\otimes N}\,,
\end{equation}
where $W_i$ is an arbitrary unparameterized unitary and $R_l(\theta_l)=\exp(-\iota\frac{\theta_l}{2}\sigma^{\alpha_l}_{n_l})$ is a parameterized rotation with Pauli operator $\sigma^{\alpha_l}_{n_l}$ acting on qubit $n_l$ and $\alpha_l\in\{x,y,z\}$. This ansatz includes hardware efficient PQCs as used within our manuscript.

As notation for our circuit, we define
\begin{equation}
	U_{[l_1:l_2]} := W_{l_2}R_{l_2} \cdots W_{l_1} R_{l_1} \, .
\end{equation}
The derivative of a PQC $\ket{ \psi(\theta)}=\ket{ \psi}$ in respect to the $l$-th index of the parameter is given by
\begin{align*}
	\partial_{l}\ket{ \psi}
		& = U_{(l:M]}  W_l  \partial_{l} R_l(\theta_l) U_{[1:l)}\ket{0} \enspace , \\
		& = U_{(l:M]}  W_l R_l(\theta_l) (- \iota \frac{\sigma_{n_l}^{\alpha_l}}{2}) U_{[1:l)}\ket{0} \enspace , \\
		& = U_{[l:M]} (- \iota \frac{\sigma_{n_l}^{\alpha_l}}{2}) U_{[1:l)}\ket{0}^{\otimes N} \,.
\end{align*}
We now discuss how to calculate the QFI with this ansatz.
The QFI \eqref{eq:QFI_sup} consists of two terms. The second term of the QFI is a product of two overlaps. Each overlap takes a simple form
\begin{equation}
\braket{\psi}{\partial_l \psi}=- \frac{\iota}{2}\bra{0}^{\otimes N}U_{[1:l)}^\dagger \sigma_{n_l}^{\alpha_l} U_{[1:l)}\ket{0}^{\otimes N}\,,
\end{equation}
which can be evaluated as a measurement of the Pauli operator $\sigma_{n_l}^{\alpha_l}$ on the quantum state $U_{[1:l)}\ket{0}$. This can be easily measured by sampling in a Pauli rotated computational basis.

\begin{figure}[htbp]
	\centering
	\includegraphics[width=0.49\textwidth]{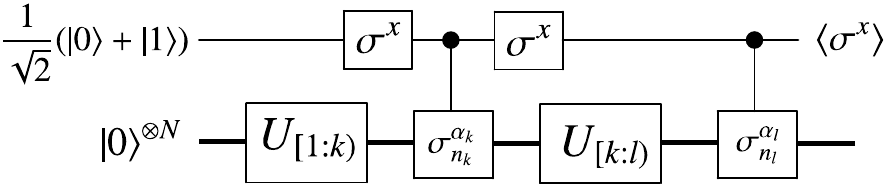}
	\caption{Hadamard test to evaluate overlap $\text{Re}[\braket{\partial_k\psi}{\partial_l \psi}]$ for QFI. }
	\label{fig:Hadamardtest}
\end{figure}

The first term of the QFI \eqref{eq:QFI_sup} consists of two derivatives. For $l\ge k$ and our ansatz, it is given by
\begin{equation}\label{eq:overlap}
\text{Re}[\braket{\partial_k\psi}{\partial_l \psi}]=\frac{1}{4}\text{Re}[\bra{0}^{\otimes N}U_{[1:k)}^\dagger \sigma_{n_k}^{\alpha_k}U_{[k:l)}^\dagger\sigma_{n_l}^{\alpha_l} U_{[1:l)}\ket{0}^{\otimes N}]\,.
\end{equation}
For ${l=k}$, this overlap is trivial to evaluate $\text{Re}[\braket{\partial_l\psi}{\partial_l \psi}]=\frac{1}{4}$.
For ${l\ne k}$, this overlap is not an observable and takes a complex number in general. 
Here, the Hadamard test can be employed~\cite{ekert2002direct,yuan2019theory,wierichs2021general}. The Hadamard test calculates overlaps of two quantum states $\braket{\psi_1}{\psi_2}$ and can measure both real and imaginary parts. 
To measure \eqref{eq:overlap}, we prepare an ancilla qubit in the state $\frac{1}{\sqrt{2}}(\ket{0}+\ket{1})$. The ancilla is entangled with the state
$\sigma_{n_l}^{\alpha_l}U_{[k:l)}\sigma_{n_k}^{\alpha_k} U_{[1:k)}\ket{0}^{\otimes N}$
by replacing the Pauli operators $\sigma_{n_l}^{\alpha_l}$ and $\sigma_{n_k}^{\alpha_k}$ with controlled unitaries. The corresponding measurement circuit is depicted in Fig.\ref{fig:Hadamardtest}.
Finally, we measure the expectation value of $\sigma^x$ of the ancilla and find 
\begin{align*}
\langle\sigma^x\rangle=&\text{Re}[\bra{0}^{\otimes N}U_{[1:k)}^\dagger \sigma_{n_k}^{\alpha_k}U_{[k:l)}^\dagger\sigma_{n_l}^{\alpha_l} U_{[1:l)}\ket{0}^{\otimes N}]=\\
&4\text{Re}[\braket{\partial_k \psi}{\partial_l \psi}]
\end{align*}
To calculate the QFI with this method, one requires $M(M-1)/2$ measurements with the Hadamard test for the terms of type $\text{Re}[\braket{\partial_k \psi}{\partial_l \psi}]$. Further, one requires $M$ measurements of Pauli strings to get terms of type $\braket{\psi}{\partial_l \psi}$.

The Hadamard test for our ansatz requires controlled unitaries applied on the ancilla and the qubit where the Pauli operator is acting on.
In case one wants to avoid implementing controlled unitaries and the ancilla, one can replace the controlled unitaries with direct measurements~\cite{mitarai2019methodology}.

}

\end{document}